# Challenges for the future of tandem photovoltaics on the path to terawatt levels: A technology review

Filipe Martinho

Department of Photonics Engineering, Technical University of Denmark, DK-4000 Roskilde, Denmark.

Corresponding author: Filipe Martinho (filim@fotonik.dtu.dk)

**Abstract**

As the photovoltaic sector approaches 1 TW in cumulative installed capacity, we provide an overview of the current challenges to achieve further technological improvements. On the raw materials side, we see no fundamental limitation to expansion in capacity of the current market technologies, even though basic estimates predict that the PV sector will become the largest consumer of Ag in the world after 2030. On the other hand, recent market data on PV costs indicates that the largest cost fraction is now infrastructure and area-related, and nearly independent of the core cell technology. Therefore, additional value adding is likely to proceed via an increase in energy yield metrics such as the power density and/or efficiency of the PV module. However, current market technologies are near their fundamental detailed balance efficiency limits. The transition to multijunction PV in tandem configurations is regarded as the most promising path to surpass this limitation and increase the power per unit area of PV modules. So far, each specific multijunction concept faces particular obstacles that have prevented their upscaling, but the field is rapidly improving. In this review work, we provide a global comparison between the different types of multijunction concepts, including III-Vs, Si-based tandems and the emergence of perovskite/Si devices. Coupled with analyses of new notable developments in the field, we discuss the challenges common to different multijunction cell architectures, and the specific challenges of each type of device, both on a cell level and on a module integration level. From the analysis, we conclude that several tandem concepts are nearing the disruption level where a breakthrough into mainstream PV is possible.

# 1 Introduction

Photovoltaics (PV) has now become a lot more than a field of research – it has become a mature industry on a global scale, with technology and market data trends changing rapidly every year. As of April 2020, the International Energy Agency estimated that 12 countries had at least 5% of their electricity generated from PV, with the top three countries reaching market penetrations of 14.8% (Honduras), 8.7% (Israel) and 8.6% (Germany), on an installed capacity basis [1]. Globally, PV currently accounts for only about 3% of all the electricity generated on Earth [1,2]. However, it is being forecasted that the global penetration of PV will increase to over 20% until 2030, and PV will likely have the largest installed capacity of any energy source after 2035 [2]. So far, this remarkable growth has essentially been enabled by two fundamental aspects: 1) the gradual improvement in the performance and operation lifetime of PV modules, resulting in a Levelized Cost of Energy (LCOE) at parity with other energy sources [3]; and 2) the success of large-scale manufacturing of PV modules, resulting in a commoditized industry with increasingly lower costs [4]. In assuming this growth of PV penetration into the future, the fundamental scientific question arises of whether it is realistic to expect both the existing and the emerging PV technologies to continuously improve, reduce costs and sustain such lofty growth forecasts. In particular, as the photovoltaic sector approaches 1 terawatt (TW) of cumulative installed capacity [4], what will be the main technological challenges for PV to achieve multi-TW levels worldwide? In this work, we review key recent trends in the photovoltaic field and discuss the main expansion limitations both on an industrial and on a fundamental research level.

Currently, one of the main fundamental barriers for that continuous improvement in commercialized devices, most notably those based on crystalline silicon (c-Si), is that their efficiency is approaching their single-junction theoretical limit, as described by the Shockley-Queisser (SQ) limit [5] or the specific Auger limit for c-Si of about 29% [6,7]. On the other hand, given the standardization of module manufacturing and the extremely low Si raw material costs, we identify the module efficiency as the key metric to improve in order to achieve further cost reductions in any future photovoltaic technology. Therefore, developing new concepts that surpass the SQ limit without a heavy increase in cost or manufacturing complexity should be regarded as one of the main challenges of current and future PV research, as discussed recently by Green and Bremner [8]. These beyond-SQ concepts include hot carrier photovoltaics [9], quantum confinement and multiple exciton generation [10], intermediate band (or impurity) photovoltaics [11], up-conversion and down-conversion photovoltaics [12,13], thermophotovoltaics [14] concentration photovoltaics [15] and multijunction photovoltaics [16]. Of these, the use of multijunction devices is by far the most well-established, as these have been the preferred choice in the space industry already since the early 1990s, based on III-V semiconductors [17–19]. In particular, the most promising multijunction configuration is based on multiple light-absorbing materials and their respective carrier-selective junctions stacked optically in series, in a tandem cascade, giving it its characteristic name of tandem photovoltaics. Although outstanding AM1.5G efficiencies of up to 39.2% have been demonstrated for tandem solar cells [15], their upscaling to TW levels has been considered unfeasible due to costs at least one order of magnitude higher than conventional single-junction devices [16]. However, recent advances in Si-based tandem cells, in particular the advent of perovskite/Si tandem devices, with a notable published record efficiency at 29.15% [20], and a new certified record at 29.52% [21], and thereby above the Auger efficiency limit for single Si, point to a possible low-cost implementation of highly-efficient tandem devices. There are several techno-economic review works available on different types of tandem devices, namely III-V tandems as used in space applications [22–26], Si-based tandems [16,27–30], organic tandems [31] and, more recently, perovskite-based tandems [29,32,33]. However, a fundamental global review on the current status and challenges of multijunction photovoltaics is needed to assess the realistic potential of this technology. In this review work, we provide a broad comparison between the different types of tandem devices, coupled with analyses of new developments in the field. We discuss the challenges common to the different tandem architectures, and the specific challenges of each type of tandem device, both on a cell level and on a module level. We highlight the fundamental functional advantages of tandem solar cell devices, and identify possible pathways to transition from single to multijunction configurations and achieve mainstream tandem devices for electricity production on Earth. The work is divided in the following sections:

**2 The scaling challenges for photovoltaics on the path to TW levels**

    Several different single-junction solar cells have now been realized nearly to their full theoretical potential. Monocrystalline Si solar cells have reached an efficiency of 26.7% [34], GaAs solar cells have achieved 29.1% efficiency [35], CuInGaSe$_2$ (CIGS) solar cells have achieved 23.4% [36] and CdTe solar cells reached 22.1% [35]. In the particular case of c-Si solar cells, remarkable advances have occurred in the industrialization of this

technology, making it a clear frontrunner in the industry. For electricity production on Earth, the market is almost completely dominated by crystalline Si (monocrystalline or multicrystalline), with a market share of above 90% consistently for decades, and climbing up to 95% recently [4,37]. The remaining 5% are currently split between the mature thin film technologies of CdTe and CIGS. The large-scale industrialization of Si module manufacturing has resulted in appreciable cost efficiencies and technological innovations, which have consistently brought down the price of photovoltaics in the recent decades. One metric used to characterize this evolution is the cost per unit power (e.g. $/W) as a function of cumulative PV module shipments, which has historically followed a trend known in the field as Swanson's law. Swanson's curve essentially shows that for every doubling of PV capacity, the price of the technology has fallen by 23.5% historically since 1976, or by 40% since 2006 [4]. This behavior is typically associated with the large-scale fabrication of disruptive technologies, and is comparable to Moore's law for electronics or, more recently, to the cost reduction curve for lithium-ion batteries, and is a subset example of the more general Wright's law of experience curves after Theodore Wright in 1936 [38].

With crystalline Si at the helm, the industry's cumulative installations have grown with a compounded annual growth rate (CAGR) of 35% since 2010 [39]. The annual evolution of installed capacity is shown side-by-side with the PV module cost reductions in **Figure 1 (a)**. For the past 10 years, the additional yearly installed capacity has grown on average by 12 GW per year. What are the implications of this growth rate for the future outlook of the PV sector? To put these numbers in context, we have compiled a direct extrapolation of this growth rate into the future decades. We assume that the 12 GW/yr growth rate in installations is maintained into the future. We do not assume any specific technological improvements, simply an expansion in capacity. To further simplify the model, we neglect the degradation and retirement of old modules from the calculations.

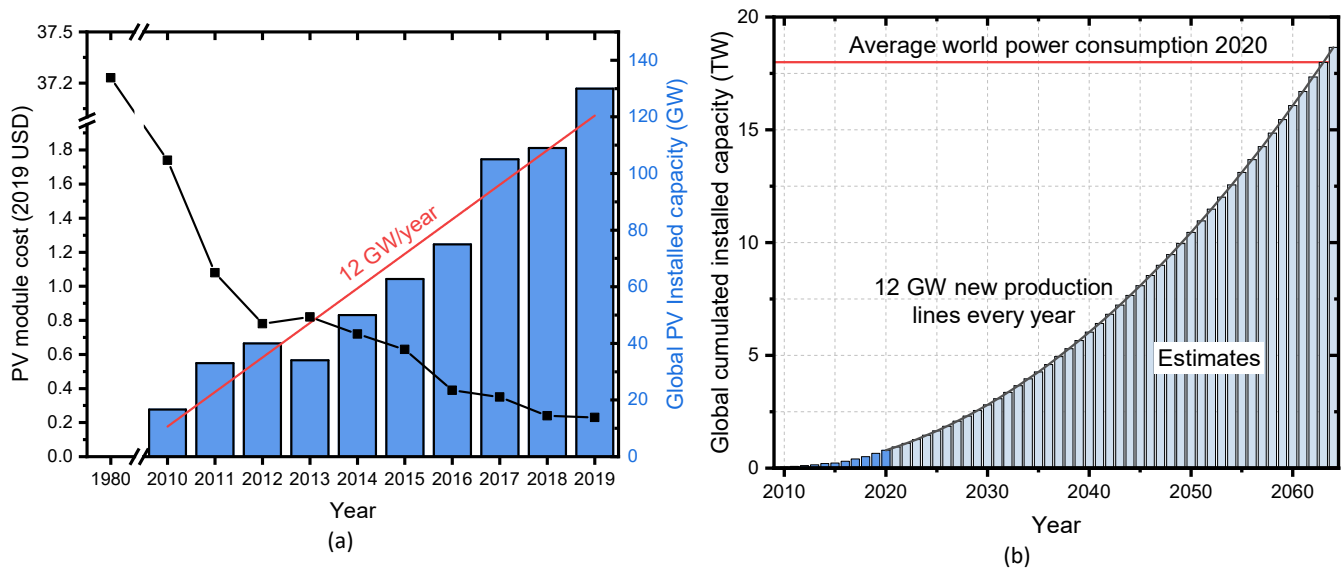

**Figure 1** – (a) Annual evolution of installed PV capacity with cost overlaid. After historical data compiled by VDMA [4] (b) Growth model linearly extrapolating the solar PV capacity assuming a CAGR of 35%, similar to 2010-2019 rates;

The 12 GW/yr growth rate in **Figure 1 (a)** can be identified as the second order derivative of the cumulative installed capacity over time. We use as initial conditions the installed power of 130 GW (2019) and the cumulative capacity of 654 GW until the end of 2019 [4]. The extrapolation is trivial, and results in the parabolic growth function plotted in **Figure 1 (b)** along with the existing data until the end of 2019. A reference line corresponding to the average world power consumption in 2020 is shown in red for comparison. The world power consumption can be estimated at around 18 TW by taking the total electricity consumption of ~30000 TWh expected in 2020 [2] and assuming a moderate effective capacity factor of about 19% across all power generation sources. Incidentally, the capacity factor of a solar PV module in a location with 1500 yearly sun hours, typical for southern European locations, is 17%. Therefore, the similar capacity factors allow for a nearly direct comparison in the energy production forecasts if, for further simplicity, we assume that the power generation and demand curves can be matched (for instance by means of storage). In that case, the model would predict that, at this 12 GW/yr rate, solar PV would reach the entire world power consumption during the 2060s. How realistic is this extrapolated model? While it certainly requires a significant long-term growth in the sector, there are several factors that make it feasible:

- Solar PV will not need to provide 100% of the energy needs, given the growth in other renewable energy sources such as wind. Considering only 50% for solar PV, the target can be achieved already in 2045.
- Considering the current manufacturing data, the growth rate of 12 GW/year is essentially equivalent to building 2 manufacturing facilities (fabs) per year (new fabs are currently >5 GW each [4]). Moreover, the processing line speed of current Si tools is at 7000 wafers/h [4]. At 5W per wafer, this corresponds to 35 kW per tool per hour. Innovations in manufacturing are expected to further increase this rate. Moreover, just CdTe alone added nearly 6 GW of new lines in 2019 [39]. However, for comparison, the *total* PV installations in the United States in 2019 amounted to only 13.3 GW [1], which indicates that a capacity expansion of 12 GW/yr should be regarded as a highly ambitious rate, implying significant efforts coordinated on a planetary scale.
- Energy storage advances will significantly increase the flexibility of PV systems and the efficiency in power consumption and distribution. In particular, the cost of lithium ion batteries has also fallen exponentially in the last decade, due to its own technology learning curve. The combination of solar PV plus battery storage at very low prices will increase the demand for solar PV [40]. A recent study suggests that the combination of wind, solar PV and energy storage could realistically provide cost-competitive baseload electricity [41]. Moreover, new applications such as agrivoltaics, floating photovoltaics, building-integrated photovoltaics such as solar roof tiles, and emerging markets in Asia and Africa could sustain further growth in demand.
- In this model, we have neglected the retirement of old modules. Considering the current data on Si module degradation of below 0.5%/year (absolute %) and warranties now at a minimum of 25 years [4], this is not expected to shift the model predictions by a large factor. In fact, in a report from June 2020, the Lawrence Berkeley National laboratory found that, in the US, the average project lifetime expected by solar PV developers was 32.5 years (up from 21.5 years in 2007) [42]. One manufacturer quotes a useful life of 40 years, where at least 99% of the modules retain at least 70% of their performance [43]. In data from a 2010 report, only 2% of modules failed after 12 years of operation [44]. As of April 2019, one other manufacturer reported that out of 4 million modules installed in Europe, a failure rate as low as 0.0044% was observed [45].

- There is no fundamental supply limitation in crystalline Si for further upscaling to meet the increasing demand. Silicon is the second most abundant element on the Earth's crust (after oxygen) [46]. The main obstacle for further upscaling of Si module capacity has been identified to be the availability and price of silver, used in the metallization of modules [47,48]. The most recent 2020 forecasts indicate that photovoltaics now consumes 9.8% of the global Ag supply [49]. Plugging these recent mid-2020 Ag supply and demand numbers into our model results in the projections of **Figure 2**. Without any innovations on Ag reduction in its use in module metallization, our model predicts that the photovoltaic sector will become the biggest single consumer of Ag after 2030, accounting for 27% of the mining volume in 2034. However, with Ag reductions as predicted by recent industry surveys [4], 2020 levels could be sustained into the future even with the parabolic growth in demand of our model. On the other hand, Ag is a very expensive metal. The current all-time high Ag price on the commodities market was $1.94/g, reached in 2011. At median 2019 levels of 100 mg/cell [4] and considering 5 W per cell (wafer), this would correspond to a price of $0.039/W just due to Ag costs alone, which would be 17% of the whole 2019 PV price of $0.23/W! Using the current December 2020 price of $0.842/g instead, this becomes $0.017/W or 7.4% of the 2019 cost of PV. Additionally, according to similar estimates, the current Ag price represents 33% of the price of non-wafer components, both in monocrystalline and in multicrystalline Si wafers [4]. This small example illustrates that crystalline Si PV is approaching the limit where the influence of the cost of raw materials becomes significant. To overcome these supply and cost limitations, several strategies can be developed, such as an increase in Ag sourcing from recycling (in particular of previous generation modules), and innovations in metallization materials and processes [4]. Note that these figures also improve if the cell efficiency increases: a 24% cell efficiency results in 6 W/cell, compared to 5 W/cell for a 20% cell, assuming a standard 158.75×158.75 mm$^2$ wafer size of existing lines. Therefore, efficiency increases of the core cell technology also mitigate Ag metallization costs. On the other hand, n-type cell concepts use a significantly higher amount of Ag, with over 150 mg/cell compared to under 100 mg/cell for p-type, due to simultaneous front and rear metallization requirements [4].

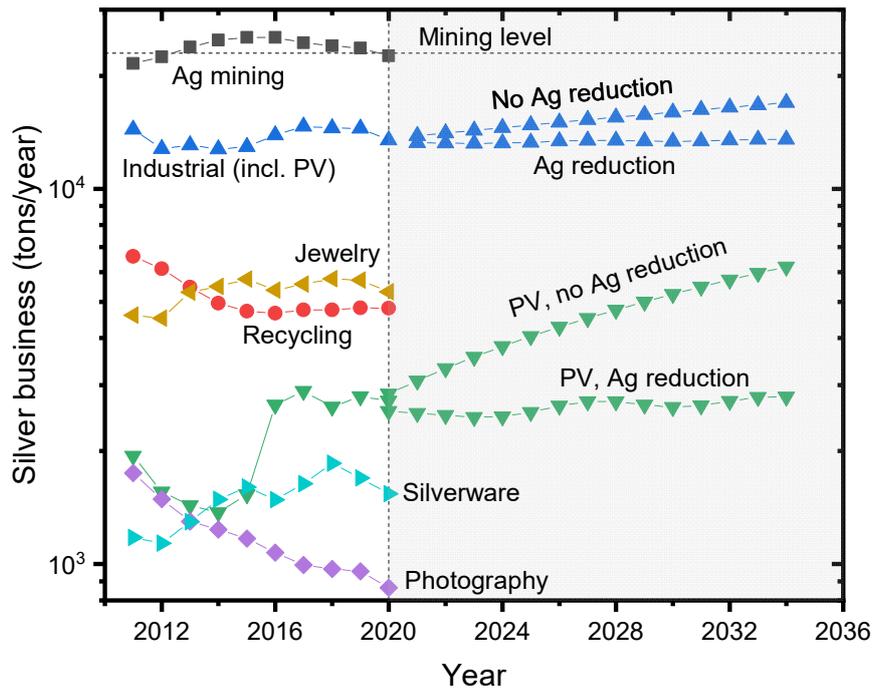

**Figure 2** – (a) Historical evolution of Ag supply and demand per sector (after data from The Silver Institute [49]), compared to the PV growth model discussed in the text and Ag reduction forecasts by PV industry surveys [4].

- Despite their lower market share of 5%, the established thin film technologies of CIGS and CdTe can potentially play an important role to help sustain this growth in PV capacity. Even with the inherent limitations of tellurium and indium scarcity, it was estimated in 2000 that the yearly growth in installed capacity would be capped by Te and In availability at 20 GW/yr for CdTe and 70 GW/yr for CIGS, respectively [50]. These estimates assumed module efficiencies of 10%, which have in the meantime improved to above 17%, as discussed later in this section. Note that even at that, these values are several times larger than the growth rate of 12 GW/yr assumed in our extrapolation model. Moreover, dedicated recycling systems could further mitigate these material constraints. Therefore, in combination with c-Si, this growth rate could in principle be sustained without critically loading the planet's resource availability.

- One factor which is mostly overlooked in photovoltaic literature is that there is often an oversupply of certain metals which are byproducts of mining of metals used on a wider scale – the so called base metals. For instance, tellurium, indium, gallium and cadmium are not mined directly, but instead are byproducts of the following base metals [50,51]:
    - Te is a byproduct of Cu and Pb mining;
    - In is a byproduct of Zn and Pb mining;
    - Ga is a byproduct of Zn and Al mining, and
    - Cd is a byproduct of Pb and Zn mining

During the metal refining process, byproduct metals for which there is an economy are recovered – a notable case is gold. Otherwise, the resulting metal slime containing Te and Cd heavy metals (among others) has to be disposed of in landfills as hazardous waste. The problem is that the amount of these metals cannot simply be reduced, as it is a byproduct contained in the ores and therefore is always correlated to the volume of mining of the base metals. In this situation, a better alternative would be to establish an economy where these metals could be encapsulated in a CIGS or CdTe module and produce electricity for 30+ years before finally being recycled or decommissioned. Based on this sustainability aspect alone, it would be highly beneficial for our society to develop any kind of low-carbon energy production technology based on such metals, even if they are scarce. Naturally, these technologies should be economically competitive and should employ such materials in an environmentally safe manner. Environmental studies of CdTe show that such a sustainable use is very much possible for that technology. Typical CdTe modules are highly resistant against leakage (or leaching) even during a fire, with a study showing that between 99.5-99.96% of the cadmium would be safely encapsulated even at temperatures up to 1050 °C [52]. Perhaps a more striking example comes from comparing it to our current energy solutions. In their whole life cycle, CdTe solar cells are estimated to emit 0.016 g of cadmium per GWh of energy produced. On the other hand, a typical coal power plant in the US equipped with state-of-the-art Cd filters will still emit 2 g of Cd per GWh [52].

Whereas further capacity expansions are not fundamentally constrained and should be expected into the future, further improvements in the core technology are now a significant challenge. Another metric that can be used to describe the remarkable improvements and maturity in the solar PV industry is the evolution of module efficiencies over time being commercialized in a given market. **Figure 3** shows the example of the US solar PV market from 2002 to 2018. There is a notable increase in the mean module efficiency over time, showing that there has been value added to the consumers even with the reduction in costs. The data also shows a progressive increase in the market share of monocrystalline Si solar cells concepts, which represented around 90% of the US market in 2018. Other publications predict that monocrystalline technology will continue to gain global market share into the future and become nearly ubiquitous [4]. This trend is occurring because the continuous decrease in the cost of Si as a raw material means that it is no longer the limiting factor, so a larger emphasis is put on module efficiency, which is notably higher in monocrystalline Si (mono-Si) cells. The increase in new mono-Si production lines reflects the maturing Passivated Emitter and Rear Cell (PERC) technology, which has allowed module efficiencies to approach and surpass 20% with low manufacturing costs [53,54].

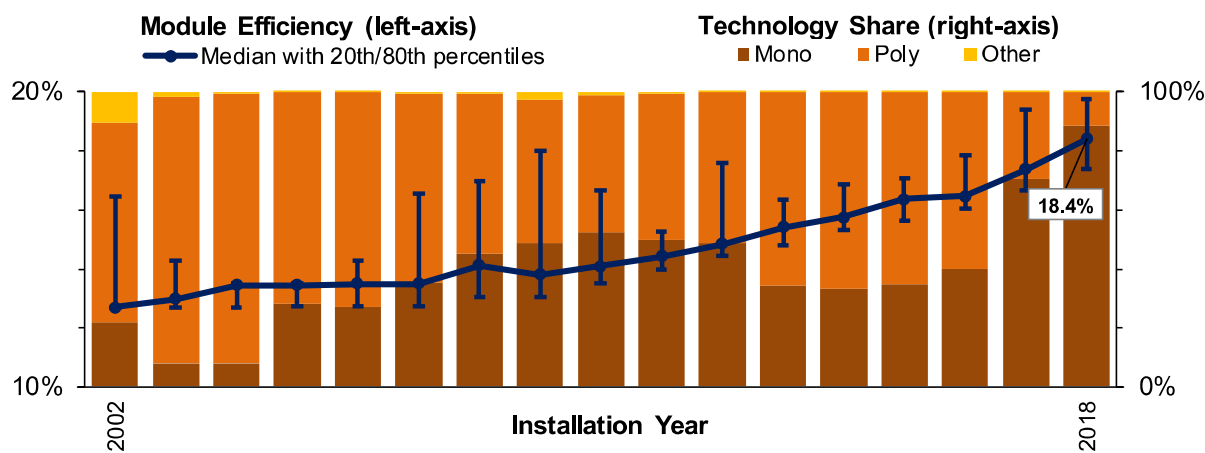

**Figure 3** – Module efficiency and technology share trends over time in the commercial market in the United States. Figure after Barbose and Darghouth [55], reproduced with permission, copyright ©2019 United States Department of Energy, under Contract No. DEAC02-05CH11231.

Perhaps one of the most notable aspects of this analysis is that in 2002, when this data series starts, the best research cell efficiencies were already around 24.8% for mono and 20.5% for multicrystalline Si (mc-Si), but the median module efficiency in the US sold that year was below 13%. This data tells the story of a maturing industry. While the cell efficiencies were already closely approaching their theoretical limit (in particular mono-Si), years of industrial development were still required to reproduce these efficiencies on solar cells manufactured on an industrial scale, and to increase the cell-to-module coupling efficiency. Today, we are at the latest stages of that industrial maturing, with the replacement of old Aluminum Back Surface Field (Al-BSF) lines, used since the mid-1980s, with modern high-efficiency concepts such as PERC, silicon heterojunction (SHJ, also known as heterojunction interfaces with intrinsic thin layers – HIT), interdigitated back contact (IBC), and polycrystalline Si on oxide passivating contacts (POLO, also known as tunnel oxide passivating contacts – TOPCon) [53]. As a result of this maturing industry, solar PV has become an extremely competitive, low-margin commoditized market, with several manufacturers competing globally with modules at the 20% efficiency level, as shown in **Figure 4**. The established thin-film technologies of CIGS and CdTe can be seen lagging behind in terms of module efficiency, with under 17.5% for CIGS, 18.2% for CdTe, followed by 18.8% for mc-Si which, as mentioned above, is being nearly pushed out of the market because their lower cost does not fully compensate their inferior performance compared to mono-Si. In fact, the technological developments in this field are progressing at such a high pace, that a couple of years are enough to significantly change this solar PV landscape. As an example, up until 2015, when commercial Si module efficiencies were at the 16% level, the success of the CIGS, CdTe and mc-Si technologies was that they offered competitive module efficiencies with much lower costs and much shorter energy payback times, as detailed in a 2016 report by Woodhouse and coworkers [56]. However, with the recent cost reductions and simultaneous efficiency advances to over 20% in mono-Si modules, the balance is shifting heavily in favor of mono-Si. This remarkable success of mono-Si is hampering further developments and leading to bankruptcies in the thin film industry. According to public disclosures, out of the biggest CIGS manufacturers, Miasolé has ceased operations since October 2019 [57], and Solibro, after announcing insolvency in August 2019, has been liquidated in March 2020, as no saving investors could be found [58]. A similar unfortunate ending seems to have come to Alta Devices, a commercial GaAs cell manufacturer who currently holds the record for the highest single-junction solar cell efficiency, of any kind, of 29.1%. Alta Devices ceased operations in early 2020 [59]. In the case of GaAs, the high costs of the metal oxide chemical vapor deposition (MOCVD) method used could not compensate for the higher efficiencies of this technology, so the use of GaAs in terrestrial application has been restricted to niche cases only. Recently, however, there have been new research developments using hydride vapor phase epitaxy (HVPE) which could lower the cost of the GaAs technology and enable large-scale terrestrial applications [60].

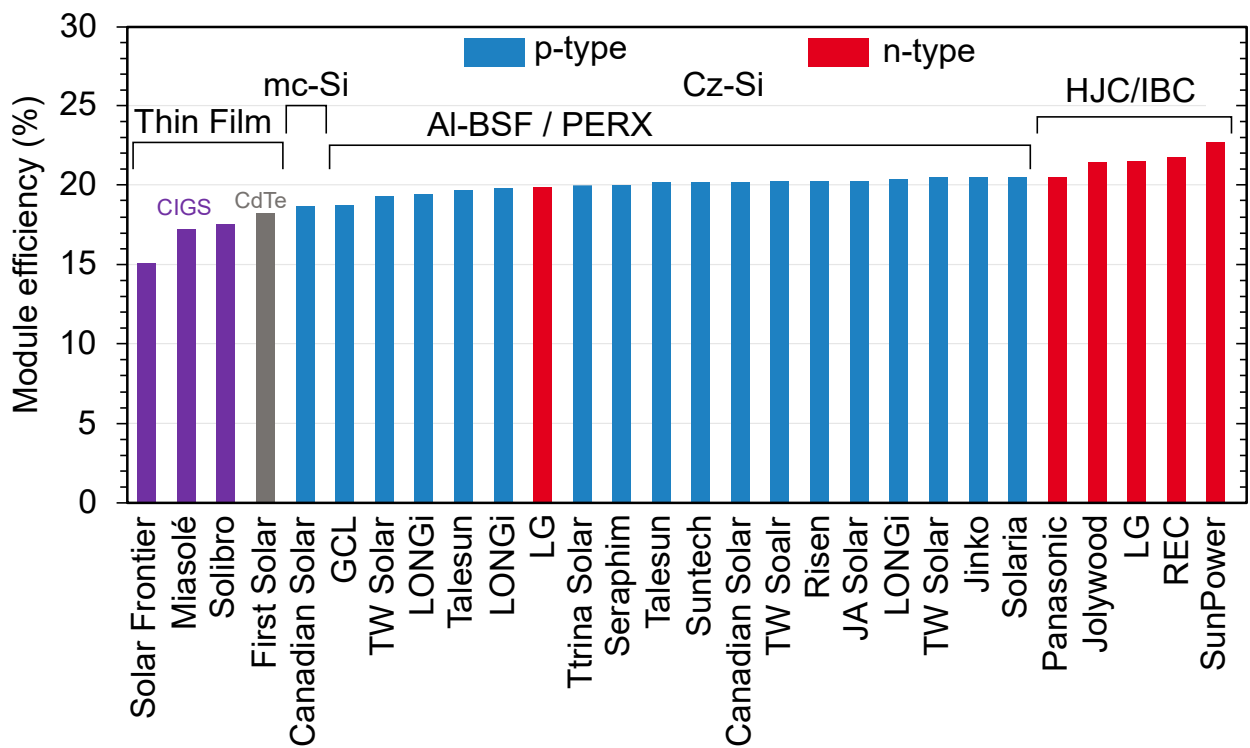

**Figure 4** – Outline of current efficiency and technology of some of the biggest vertically-integrated solar PV manufacturers. Data based on company product data sheets to the best of the author's knowledge, modified with permission from ©Fraunhofer ISE: Photovoltaics Report, updated: 16 September 2020 [37]. Disclaimer: the figure should not be considered a complete list of manufacturers and product specifications, and is not intended for commercial purposes.

These trends paint a new picture for photovoltaics into 2021, and will potentially shape the directions of future photovoltaic research and development in the industry and in academia. Due to the economies of scale and the developments in the crystalline Si industry surpassing even the most optimistic forecasts, the cost of producing a PV module now represents less than 40% of the overall cost per watt of PV. The other 60% represents the system cost, also referred to as the Balance of System (BOS) cost, as detailed in **Figure 5**. These BOS costs are more or less technology agnostic if one reasonably assumes that there is no major disruption in installation configurations from one technology to the other. Then, of that 40%, more than 50% of the cost is related to non-silicon material costs such as consumables and materials used in module assembly, such as encapsulation materials, back sheet materials, soldering, framing and metallization materials, of which Ag is an example, as mentioned above [4]. Although some parts of this module cost breakdown are technology-dependent – for instance monolithic modules vs. wafer-based modules –, the need for protection, long-term durability and reliability (warranty) standards means that we can expect similar costs for modules of any technology [8]. This also has the positive effect that any new technology will likely benefit from these same BOS cost decline curves.

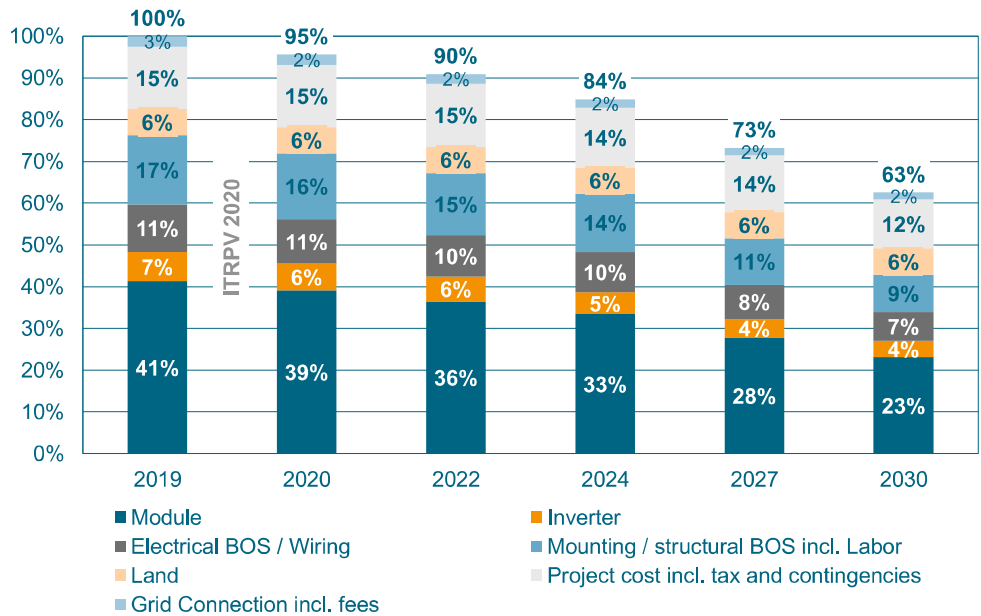

**Figure 5** – Cost breakdown of PV systems worldwide, for utility-scale projects (>100 kW). Figure reproduced with permission from [4], copyright ©2020 VDMA.

Furthermore, since the majority of the BOS costs are infrastructure-related, they tend to scale with the project area (one exception being the inverter cost). This means that lower efficiency modules are not favorable because they will imply larger areas and infrastructure costs, which now represent the biggest fraction of the overall costs. Likewise, this also means that even if a new breakthrough material with extremely low fabrication costs and efficiency comparable to mono-Si is developed, it would only lower the overall cost of PV by about 20%. This is exactly what is happening currently with the developments of organometal halide perovskite solar cells. The reader who attempts to tackle the exponentially growing literature on perovskites will quickly realize that perovskite solar cells have the potential to be the high-efficiency photovoltaic technology with the lowest cost, by far – as long as the very expensive spiro-OMeTAD hole transporting material and the use of gold contacts are avoided, which has been successfully demonstrated recently [33,61]. In a recent work, Rolston et al. claim that large-area spray-coated perovskite modules with 15.5% efficiency could achieve the lowest cost of any solar technology [62]. And yet, due to this BOS cost balance, even if 20%+ efficiency perovskite modules with 25+ years lifetime and near-zero material costs are demonstrated in the near future, such a breakthrough would not move the needle more than 20% on the *current* costs of PV (other studies have found this value to be 25% or below [8]). Quite paradoxically then, even this undoubtedly impressive breakthrough would likely not add much value to the PV sector! One possible aspect in which this hypothetical breakthrough could be beneficial would be if the temporal throughput of this type of technology was significantly higher than that of crystalline Si. That is, if the power output per unit time (e.g. W/h) of the manufacturing tools was inherently higher, for instance by means of a roll-to-roll or a similar high-throughput processing method. Then, such a breakthrough technology could be scaled up and quickly provide several GW worth of modules to help society transition to a renewable energy economy and meet the global carbon budget. In principle, this would be true, but the data shows otherwise. The current data on PV expansion capacity shows that supply is far outpacing demand, with the predicted 2020 demand at 142 GW and a supply level already at 200 GW in 2019, with several capacity expansion projects underway [4]. To make matters worse, due to the SARS-CoV-2 Covid-19 pandemic, the real 2020 and 2021 demand numbers will be below forecasts (although final 2020 numbers are not yet available at the time of writing, a demand contraction of roughly 15% can already be predicted for 2020 from quarterly financial disclosures of inverter and module sales from the major manufacturers [63]). This large supply and demand unbalance is one of the reasons why the costs of PV are being kept low, and essentially shows that we, as a global collective society, are adopting solar PV far too slowly, even slower than the expansion of capacity, despite the currently very competitive costs of solar PV. This now gives our extrapolation model another perspective: the assumed growth of rate of 12 GW/yr is actually being constrained by lack of demand, not by the possible objective scientific challenges discussed above. Further demand will be unlocked by continuous reductions in LCOE for PV modules and projects, and only in that case should high throughput technologies become relevant.

The previous discussion indirectly highlights what is now, more than ever, the most important metric in research to achieve future improvements and increase the value added by PV: the module efficiency. As we have seen from **Figure 5**, the majority of the costs of PV are now infrastructure or area-related. Therefore, whereas a 20% material cost reduction only results in a maximum of 4% reduction in the $/W cost of PV, a 20% increase in the module efficiency directly results in nearly a 20% decrease in the cost of PV, as long as that increase in efficiency does not significantly add to the module manufacturing and material costs. In accordance, the recent PV market trends show a shift towards higher efficiency cell types and towards concepts that increase the module energy yield. The latter includes for instance solar tracking or bifacial cells. In particular, bifacial Si cells have reached a 20% market share in 2019, and are expected to grow to a notable 70% share by 2030 [4]. Nevertheless, further efficiency (or energy yield) increases will soon be intrinsically limited: the current commercial technologies of Si, CdTe, CIGS, are based on solar cells with a single p-n junction, and are therefore limited to an efficiency of ~30% by the Shockley-Queisser limit, mentioned in the beginning of this section. In this context, the transition to multijunction photovoltaics has the potential to provide a major step improvement in module efficiencies. In particular, the first and simplest transition, from single-junction to double-junction cells concepts, could result in up to 47% relative increase in cell efficiency when using Si as a bottom cell [8].

# 3 The transition to multijunction photovoltaics

The progression from single-junction to multijunction photovoltaics occurred early in the 1990s for space applications, using III-V semiconductors. We will briefly review this transition process as it describes the key merits of each photovoltaic technology and reveals the fundamental advantages of tandem cells. Then, we will review the different tandem fabrication strategies, their potential and challenges. Following this, we will see how the different strategies can result in cells with different electrical configurations, and how that can impact the device performance.

*3.1 Space photovoltaics: the functional advantages of tandem devices*

Historically, there have been many technology exchanges between the terrestrial and spatial sectors, and these continue until today. In particular, most of the advances in the early days of the modern photovoltaic era were due to space exploration. Solar cells were used as early as in the 1958 Vanguard 1 satellite, the fourth satellite launched after the Sputnik 1 and 2 (1957) and the Explorer 1 (1958). Vanguard 1 was powered by six 0.5 x 2 cm$^2$ Si solar cells, offering a power of a few mW to the satellite [19]. However, in 1959 it was discovered that Earth was surrounded by ionizing radiation belts, later named Van Allen radiation belts, which consist of two belts with a flux of high-energy electrons and protons. Most other planets in our solar system exhibit similar radiation belts, except Venus and Mars, which do not have core magnetic fields. These radiation belts can cause severe damage to the satellites' solar cells and microelectronic circuits. In the case of solar cells, this radiation significantly reduces the minority carrier lifetimes and the quality of the interfaces by creating lattice defects [64]. Silicon-based devices were particularly shown to degrade, leading to questions whether solar cells would be useful in space at all [19]. In 1962, it was predicted that GaAs would suffer less from radiation damage than crystalline Si, which motivated intense research in GaAs and eventually other III-V semiconductors [19]. By the 1980s, single-junction III-V GaAs started being preferred for use in space, even though at that time it was already 5-10 times more costly than crystalline Si [18,23]. During the 1990s, Si space solar cells fell out of use due to the advances in III-V tandem concepts, demonstrating higher starting efficiencies, and the discovery of the high radiation resistance of InP, AlGaAs and InGaP alloys [17–19]. This led to the development of efficient multijunction III-V tandem devices, throughout the 1990s, in particular GaInP/GaAs double-junction (2J) and GaInP/GaAs/Ge triple-junction (3J) devices, which achieved at that time 21-22% and 24% efficiency, respectively, in a volume production setting, and up to 30% in small-scale research devices [23,24]. For that reason, space applications started using III-V tandem devices almost exclusively since the late 1990s. In particular, the $Ga_{0.5}In_{0.5}P/Ga_{0.99}In_{0.01}As/Ge$ triple-junction solar cell has been the most commercially successful solar cell of any kind in the space industry [65]. This kind of solar cell has reached a one-sun AM1.5G efficiency of 32.0% [66]. **Figure 6** illustrates the fundamental principle allowing the higher efficiency of multijunction photovoltaics, using the GaInP/GaInAs/Ge triple-junction tandem cell as an example. The principle consists in combining different absorber materials to match a certain part of the solar spectrum. A single absorber cannot absorb the fraction of solar photons below its bandgap, resulting in electrical current losses. On the other hand, photons with energy much higher than the bandgap will result in hot carriers that quickly relax to the band edges, within $10^{-12}$ s [67], by interaction with the crystal lattice, resulting in voltage losses. Therefore, a set of multiple absorbers with different bandgaps can more efficiently convert the solar photons into electricity. For a theoretical description of these improvements in tandem devices, the reader can refer for instance to the work by Hirst and Ekins-Daukes [68] and references therein. Instead, in this review we focus on the engineering challenges associated with fabricating these devices. The corresponding device essentially consists of a cascade of absorbers with decreasing bandgap, each with the necessary carrier membranes and with suitable interfaces between each subcell. The non-absorbed radiation, with higher wavelengths, is transmitted into the next subcell.

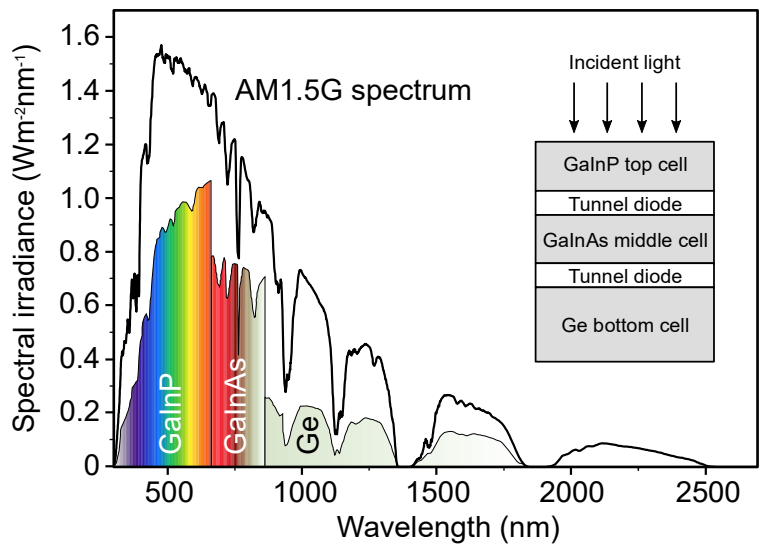

**Figure 6** – Classical GaInP/GaInAs/Ge multijunction solar cell technology used in space applications and corresponding use of the solar spectrum. Note: The spectral interfaces between each subcell (defined by their bandgap absorption onsets) were exaggerated for illustration purposes. Figure adapted from [65] with permission, © 2018 Elsevier Ltd.

In the late 1990s/early 2000s, several data on solar cell degradation in orbit were collected by satellite missions such as the Equator-S (1997) and the COMETS [69]. **Figure 7** shows some of that data for a GaAs/Ge tandem and for single junctions of GaAs, InP, Si and CIGS. All solar cells suffered significant degradation for altitudes between 1000-20000 km. This is associated with the Van Allen proton belt, whose flux is higher in this altitude region. The data shows the advantage of the higher starting efficiencies of tandem devices, and the good radiation resistance of InP and of the polycrystalline thin film CIGS, in particular when considering the degradation relative to the initial performance.

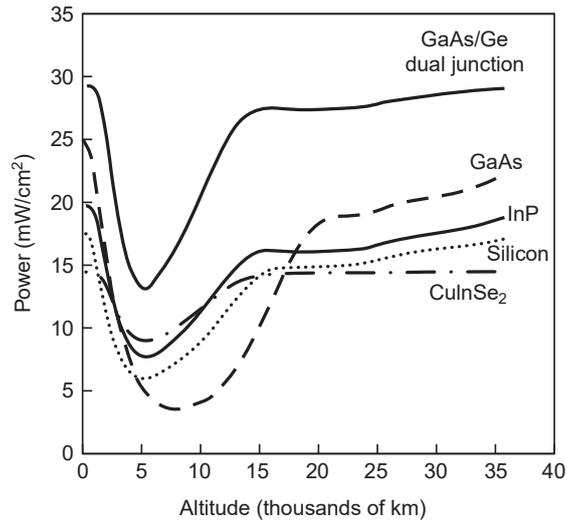

(a)

**Figure 7** – Power versus altitude measured over 10 years of satellite data from the Equator-S and COMETS missions. Figure adapted from [69] with permission, © 2018 Elsevier Ltd;

In general, the radiation-induced damage can depend on many factors, such as externally in the module coverglass thickness and type of radiation [19,70,71], and internally in the solar cell architecture, the p-n junction depth and width, the dopant concentration and even the type of dopant [17,19,64,72,73]. These internal factors are essentially related to the specific kind of radiation defects created, and where they are located (i.e. bulk, depletion region or interfaces). Fundamentally, direct bandgap materials such as GaAs tend to have a higher radiation resistance. With a direct bandgap and corresponding high absorption coefficient, all the radiation is absorbed in GaAs within a few µm of depth, meaning that the active region is also confined to a few µm near the surface, compared to around a hundred µm for Si. That makes GaAs most sensitive to particle radiation in the low energy range (250 keV to 1 MeV), which affects mostly the surface region, whereas Si will be affected by a larger energy range due to bulk lifetime degradation [73]. Similarly, other direct bandgap III-V alloys also exhibit higher radiation hardness, making III-V multijunction tandem devices suitable for high-radiation space missions.

Besides durability, space applications require the highest possible specific power (i.e., power per unit area, volume and weight). This can ultimately offset the material and production costs, because smaller array areas would allow a larger number of satellites to be launched in a single rocket, thereby saving operational and fuel costs [23,24]. In particular, the current tandem devices used in most satellites offer a specific mass of 2.6 kg/m$^2$ on an array level, but this value can be reduced to 0.6 kg/m$^2$ by transitioning from wafer-based to flexible substrates [64], as will be explained later in this section. Durable and powerful solar modules can then ensure a higher operating lifetime, further reducing the project costs. With the recent advent of the privatization of space exploration, it is expectable that the cost of PV will become an increasingly important factor in space as well. A notable example of this is the ongoing Starlink commercial mission from SpaceX, where crystalline Si cells are actually preferred over III-V multijunction cells for the satellite solar arrays due to superior cost metrics, despite the poorer performance of Si in space, as discussed throughout this section. Similarly, future commercial developments in the space industry could cause a surging interest in new low-cost tandem concepts, just like in the terrestrial case. For instance, the excellent radiation hardness of CIGS is now causing a surge in interest for CIGS/Perovskite tandems for low-cost space applications [74].

Finally, another very important fundamental advantage of tandem concepts is their lower temperature coefficient, which describes the change in efficiency $\eta$ with temperature $T$, usually normalized by efficiency in the form

$$\text{Temperature coefficient (K}^{-1}\text{ or }°\text{C}^{-1}) = \frac{1}{\eta}\frac{d\eta}{dT} \quad (1)$$

In all PV devices, the performance will degrade with increasing temperature due to an intrinsic increase in loss processes such as recombination. However, this is true even in ideal devices operating near the SQ limit, as all the absorption-emission balance losses (i.e. luminescence emission, Boltzmann solid angle mismatch and Carnot losses) increase with increasing absorber temperature [75]. As these losses depend on the bandgap of the absorber, either directly or through the current at the maximum power point, the temperature coefficient will increase (become less negative) as the bandgap increases, as shown in **Figure 8**.

**Table 1** – Temperature coefficient for selected single and multijunction solar cells, for a given efficiency and measured in a given temperature range.

| Cell type | T (°C) | $\eta$ (%) | $1/\eta\ d\eta/dT$ ($10^{-3}$ K$^{-1}$) |
|---|---|---|---|
| Best Si module 2020 [76] | -40 - 85 | 22.2 | -2.9 |
| Best CdTe module 2020 [77] | 25 - 75 | 18.2 | -3.2 |
| GaAs 1985 [78] | 0 - 80 | 16.4 | -2.0 |
| InP 1991 [78] | 0 - 150 | 19.5 | -1.59 |
| 2J GaInP/GaAs [79] | n. a. | 23.0 | -2.1 |
| 3J GaInP/GaAs/Ge [79] | n. a. | 26.0 | -2.0 |

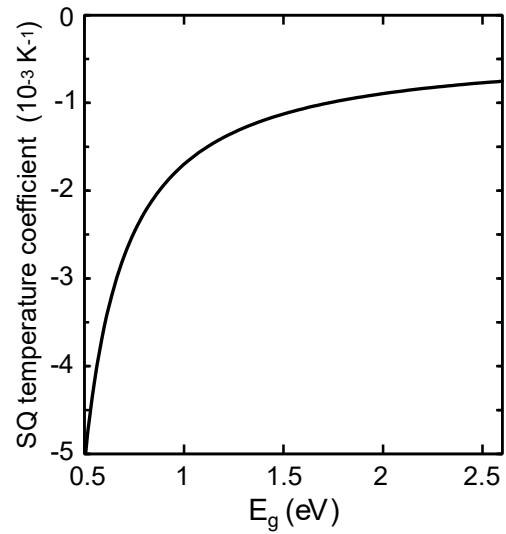

**Figure 8** – Ideal SQ-limit temperature coefficient curve. Adapted with from [75] with permission, © 2018 Elsevier Ltd.

That means that higher bandgap materials are intrinsically less temperature sensitive, both in single-junction devices, but also in tandem devices, where the higher bandgap materials in the top cells convert a large fraction of the energy. Some examples of temperature coefficients are given in **Table 1**. As the efficiencies increase with technology development, the temperature coefficients will converge to their SQ ideal limit. This lower temperature sensitivity of higher bandgap materials is a significant advantage in operations requiring high temperatures, as is often the case in space missions. For example, the illuminated surface of the Moon can reach 130 °C [22]. Taking the data for the best Si module in 2020 and for the triple-junction InGaP/GaAs/Ge, we can compare the efficiency losses as shown in **Table 2**:

**Table 2** – Calculation comparing relative degradation and resulting absolute efficiency for a 22.2% efficiency Si module [80] and a triple-junction III-V based tandem [79].

| Moon (130 °C) | Best Si module 2020 | 3J InGaP/GaAs/Ge |
|---|---|---|
| Degrad. % | $-2.9 \times 10^{-3} \times (130 - 25) = 30.45\%$ | $-2.0 \times 10^{-3} \times (130 - 25) = 21\%$ |
| New eff. $\eta$ | $\eta(130\ °C) = 22.2 \times (1 - 0.3045) = 15.4\%$ | $\eta(130\ °C) = 26 \times (1 - 0.21) = 20.5\%$ |

For these very high-temperature applications, the lower temperature coefficient is a critical advantage and can potentially outweigh the type of technology used. For terrestrial applications, average operating temperatures are typically around 44 °C [81], but still high operating temperatures up to 65 °C can be reached, for example in modules operating in desert-like conditions. In that case, the degradation rates are 11.6% for Si vs 8% for the 3J tandem. Given the importance of efficiency for the overall PV system cost, this difference could play a role if any tandem technology approaches the costs of Si.

*3.2 Tandem cell fabrication concepts: prospects and challenges*

After describing the fundamental advantages of tandem devices on more general grounds, we now proceed to analyze the technical aspects and challenges of the different tandem technologies at the cell level. We analyze the most successful tandem examples used in space applications, based on the epitaxial growth of III-V semiconductors, in lattice-matched or lattice-mismatched (metamorphic) configurations. Then, we analyze the prospects of combining III-Vs with Si to achieve high-efficiency concepts at lower costs, by means of metamorphic structures, wafer bonding or mechanically stacked cells. Finally, we analyze emerging Si-based tandem concepts, where new top cell materials such as perovskites can be combined non-epitaxially with Si by means of developing specific subcell interconnections.

*3.2.1 Lattice-matched tandem cells*

The classical example of a lattice-matched tandem cell is the GaInP/GaInAs/Ge 3J space cell, shown in the schematic diagram of **Figure 9 (a)**. Starting from a single-crystal Ge substrate, all the subsequent layers are grown in very close lattice matching with the preceding layers (<1% mismatch), forming epitaxial thin films with a very low concentration of detrimental defects and with no grain boundaries, essentially similar to a single crystal, and hence with very good photovoltaic properties. This can be achieved thanks to the very wide alloying stability of the III-V compounds, which allows a precise tuning of the lattice parameter and bandgap of the films, as shown in **Figure 9 (b)**. This effect is of significant technological importance, as it is also used in light-emitting diodes and in the low dimensional quantum structures used in solid-state diode lasers [82]. Besides the purpose of lattice matching, these III-V alloys also play a key role in enabling a lossless interconnection between each junction. This interconnection is, in fact, one of the most important challenges in the engineering of tandem cells of any kind, because the coupling losses between each subcell have to be minimized in order to achieve the maximum potential of multijunction configurations. In the multijunction configuration of **Figure 9 (a)**, each subcell consists of a p-i-n structure with a top n-doped emitter, an intrinsic (undoped) layer and a bottom p-doped base layer. If each p-i-n subcell was simply connected to the next subcell, the interfaces would form reverse diodes blocking the flow of current. Therefore, additional intermediate structures are required.

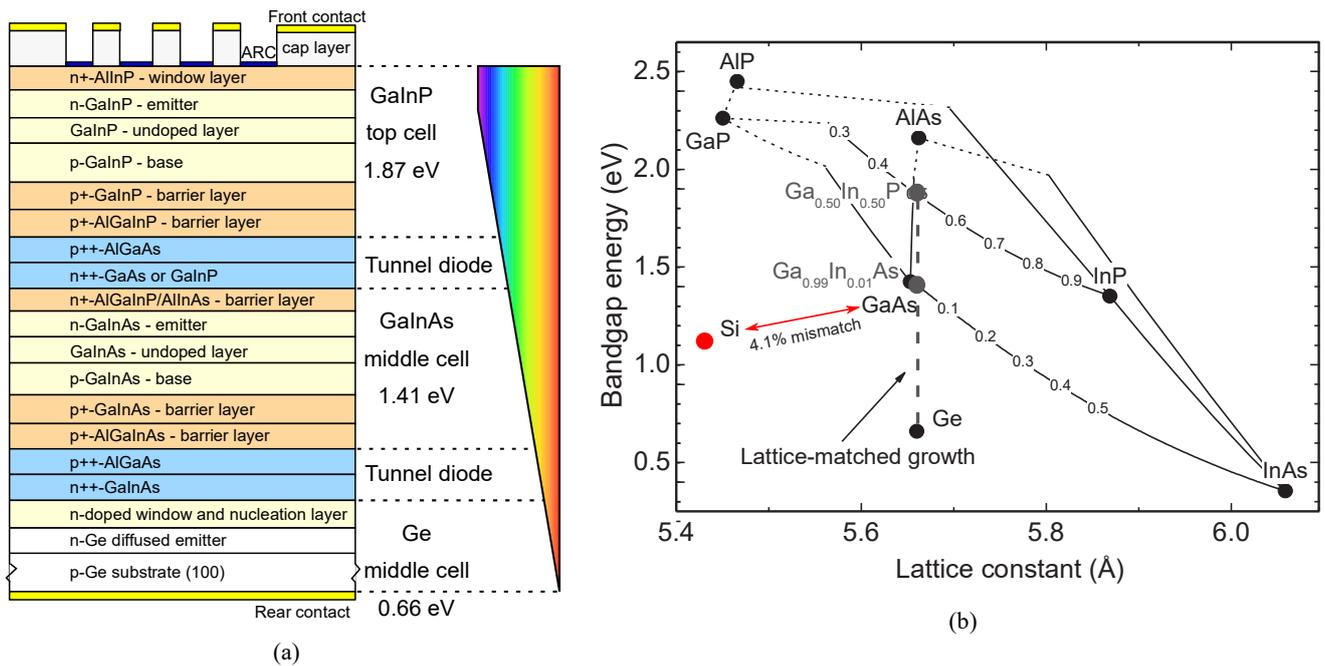

**Figure 9** – (a) Layer diagram of the monolithic lattice-matched triple-junction tandem cell based on $Ga_{0.5}In_{0.5}P/Ga_{0.99}In_{0.01}As/Ge$, widely used in space applications; (b) Bandgap as a function of the lattice constant for the corresponding III-V compounds. Ternary compounds are joined by a line, with solid lines for direct bandgap semiconductors and broken lines for indirect bandgaps. Si is included as a red point for comparison. Both figures are adapted from [65] with permission, © 2018 Elsevier Ltd.

*3.2.2 Subcell interconnection by tunnel diodes*

In these types of III-V tandem cells, the extrinsic doping of each specific layer can be easily fine-tuned during growth, or using ion implantation and diffusion processes similar to those developed for Si microelectronics and photovoltaics [83]. For that reason, the subcell interconnection is achieved by means of a tunnel junction, also called *tunnel diode* or *Esaki diode*, after Leo Esaki who was awarded the Nobel prize in physics in 1973 (shared with Ivar Giaever and Brian David Josephson) for the discovery of quantum mechanical tunneling in Ge p-n junctions [84]. This structure consists of heavily doped $n^{++}$-$p^{++}$ junctions with the same polarity as each subcell. Being heavily doped, the junction is extremely narrow, with depletion widths of just a few nm [85]. Furthermore, the semiconductors forming the junction are in fact degenerate, meaning that under equilibrium (without applied voltage and in the dark), the Fermi level lies inside the conduction band of the $n^{++}$ layer and inside the valence band of the $p^{++}$ layer. If a small positive voltage is applied, filled states on the $n^{++}$ side get aligned at the same energy level as empty states on the $p^{++}$ side, and the carriers can tunnel interband through the nm-wide narrow energy barrier defined by the p-n junction band bending. This interband tunneling can occur either directly or assisted by trap states, as illustrated in processes (1) and (2), respectively, in **Figure 10 (a)**. This nonlocal tunneling effect allows a high current density in the forward direction for voltages much lower than a conventional diode. Progressively increasing the voltage leads to an increase in the energy misalignment between the available states for tunneling, causing a decrease in the current density, effectively leading to a negative resistance region. With a further increase in voltage, the tunnel junction starts behaving as a conventional diode, as the flattening bands lead to the typical exponential diode diffusion current [65,85–87].

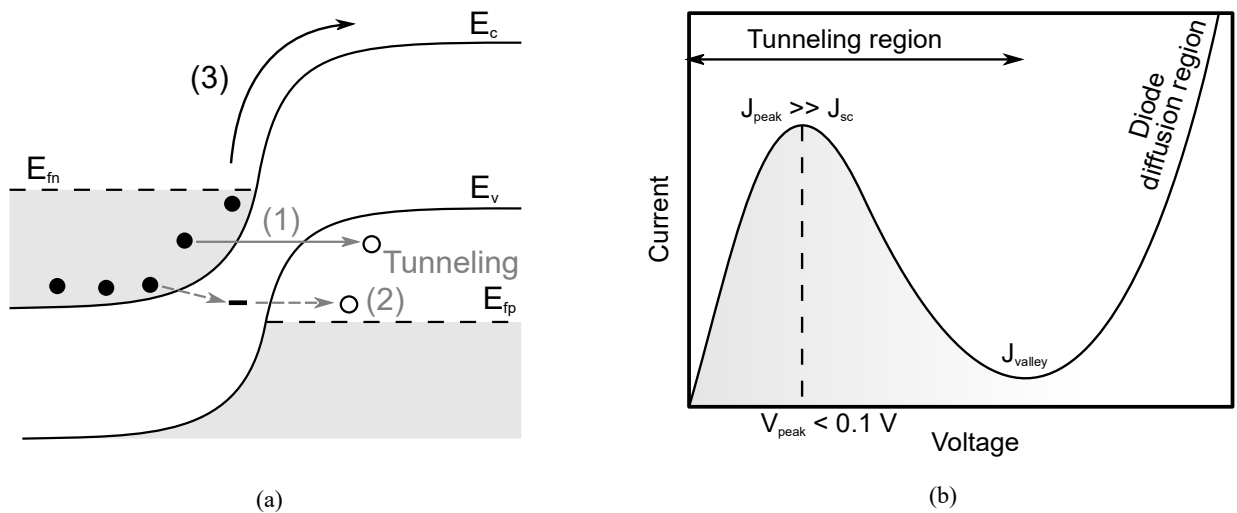

**Figure 10** – Tunnel junction band diagram showing the most important processes: (1) direct interband tunneling, (2) trap-assisted interband tunneling, (3) conventional diode diffusion current. (b) I-V characteristic curve of a typical solar cell tunnel diode.

By allowing this current injection at very low voltages, the tunnel diode creates a very low resistance path for the recombination of the *inner* carriers within each subcell interface, leaving the outer carriers to contribute to the current density of the device. Therefore, the desirable condition is for the peak tunneling current density to be much higher than the short circuit current density of the device, $J_{sc}$,

Tunnel diode condition: $\quad J_{peak} \gg J_{sc}$ (2)

at the lowest possible forward voltage. Note that the diode polarity is still reversed compared to the diode of each subcell, but now this tunneling effect allows a current injection at very low voltages, effectively acting similarly to a metal. In order to achieve the tunnel diode condition (2), the junction should be as narrow as possible, requiring doping levels on the order of $10^{19}$ - $10^{20}$ cm$^{-3}$. On the other hand, the tunneling current decreases exponentially with increasing bandgap [88], which causes transparency issues to the bottom cells. For that reason, the n$^{++}$ and p$^{++}$ layers forming the tunnel junction need to be kept as thin as possible, on the order of tens of nm. This creates an additional problem: it is very hard to highly dope such narrow regions and confine this high concentration of dopants in these thin layers during the growth of the subsequent layers. A dopant out-diffusion can cause loss of functionality in the tunnel diode. To solve this issue, additional barrier layers are used to preserve the tunnel junctions [65,88]. We have now achieved a full qualitative understanding of the GaInP/GaInAs/Ge multijunction structure in **Figure 9 (a)**.

*3.2.3 Upright metamorphic (lattice-mismatched) tandem cells*

Unfortunately, our universe was designed in a way that the bandgaps corresponding to this lattice-matched Ga$_{0.5}$In$_{0.5}$P/Ga$_{0.99}$In$_{0.01}$As/Ge device do not perfectly match the spectrum of our sun. This is due to the Ge bandgap being too low, which causes an excess current in the Ge bottom cell. Since the subcells are electrically in series, this excess current is wasted, and the tandem cell does not achieve the highest possible efficiency. This limitation cannot be easily solved by simply tuning the bandgaps with a lattice mismatch, because mismatches as small as 1% significantly increase the concentration of threading and misfit dislocations, which propagate across the subcells and degrade the overall device performance [26,65]. Therefore, additional strategies are required to achieve the full potential of this structure. One possible strategy is to introduce buffering structures with an increasingly higher lattice parameter, by using a sequence of compositionally graded III-V layers, terminated by an overshooting layer where the induced strains are relaxed [26,89,90]. With this strategy, the majority of the threading and misfit dislocations accumulate in the buffer region, providing a defect-free growth for the active layers. This effect is illustrated in **Figure 11 (a)** and **(b)**. This type of strategy is usually labelled upright lattice-mismatched or upright metamorphic (UMM) growth, as it proceeds upright from a single substrate [65].

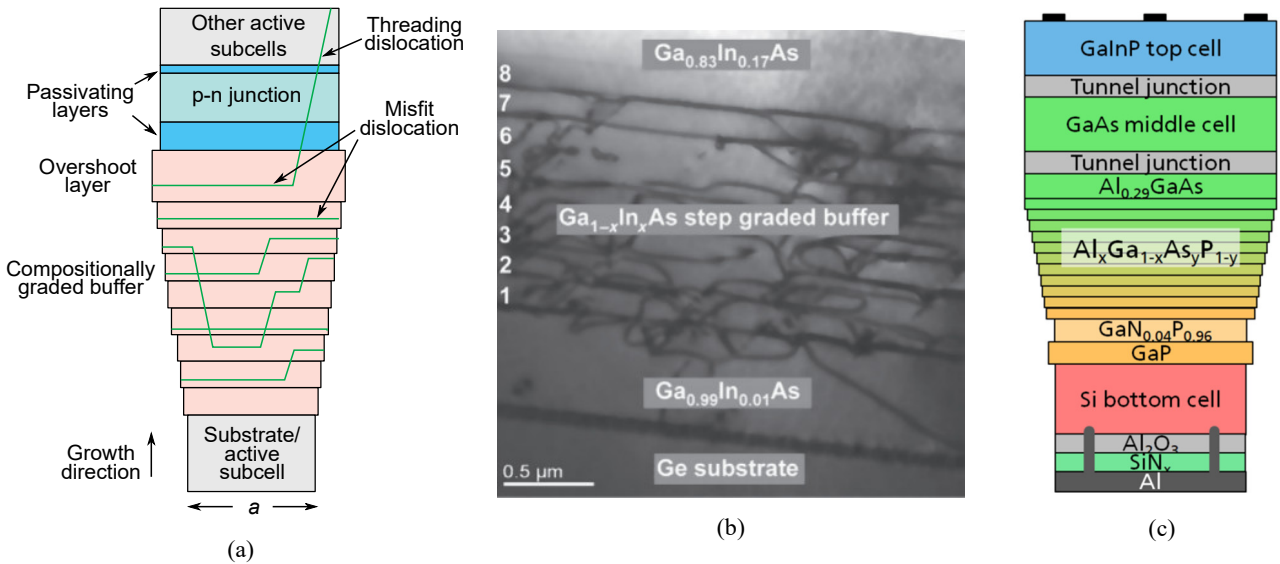

**Figure 11** – (a) Illustration of a compositionally graded buffer region allowing the metamorphic growth of tandem cells with restriction of misfit and threading dislocations in the buffer. Figure adapted from [26]; (b) Reflection electron microscopy image of the graded buffer region showing the dislocation defects. Figure reproduced from [90] with permission, copyright © 2009 American Institute of Physics; (c) General device structure of a monolithic, upright metamorphic, triple-junction GaInP/GaAs/Si tandem with an efficiency of 25.9%. Figure reproduced from [91] with permission, copyright © Fraunhofer ISE.

This upright metamorphic growth strategy can be extended to 4J, 5J and even 6J tandems, and has also achieved terrestrial efficiencies above 30% and efficiencies under concentrated light of over 41% [66,90]. It has also now been transferred to industrial production [26,65]. Moreover, this strategy could also be very relevant to allow a technology transfer into Si-based tandem cells. On one hand, it can be used to monolithically combine a Si bottom cell with III-V top cells, which would otherwise be impossible due to the 4.1% lattice mismatch between Si and GaAs, as identified in **Figure 9 (b)**. This is currently one of the most promising routes towards low-cost Si-based tandem photovoltaics, with a new 23.4% record efficiency for a monolithic $Ga_{0.75}AsP_{0.25}$/Si tandem of this kind published in June 2020 [92,93]. Using a similar graded buffer strategy between Si and III-V top cells, a 3J GaInP/GaAs/Si metamorphic tandem with an efficiency of 25.9% has been announced in August 2020 [91]. A schematic structure of this device is shown in **Figure 11 (c)**.

On the other hand, this technique could be used to implement new tandem concepts by enabling the growth of new and unexplored compounds on Si, epitaxially and free of grain boundaries. Here, chalcogenide materials such as CIGS or the kesterite $Cu_2ZnSn(S,Se)_4$ and their different alloys could be of potential interest. Being itself an alloy with III-V elements, CIGS can be combined with III-V structures, and there have been a few studies showing epitaxial growth of CIGS on III-V compounds, as reviewed by Nishinaga et al. [94]. In particular, by tuning the Ga/(Ga+In) ratio in CIGS, a precise lattice matching can be achieved with GaAs, and a dislocation free single crystal CIGS solar cell achieving 15.7% efficiency could be demonstrated [94]. By introducing a Ga grading, the efficiency could be increased to 20.0%, but this came at a cost of a lattice mismatch with GaAs, leading to dislocation defects in the CIGS absorber. The two cases are shown in **Figure 12 (a)** and **(b)**, respectively, together with the bandgap-lattice constant relation for CIGS alloys in **Figure 12 (c)**. The main challenge here is that the In content of these CIGS films was quite high, yielding absorbers with bandgaps around 1.2-1.4 eV. To combine this technology with Si, the In-poor higher bandgap alloys need to be used instead. Unfortunately, the higher bandgap alloys in the CIGS system show the poorest photovoltaic performance, as reviewed by Ramanujam et al. [95]. However, the epitaxial growth of these structures could enable significant improvements, for instance by removing the negative influence of the grain boundaries, which are notably more detrimental in the high bandgap $CuGaSe_2$ compared to $CuInSe_2$ [96,97]. A similar possibility exists for $CuInGaS_2$, which can be lattice matched to Si by tuning the In/Ga ratio, and is currently the best performing high-bandgap chalcopyrite material known, having achieved a single junction efficiency of 15.5% [98].

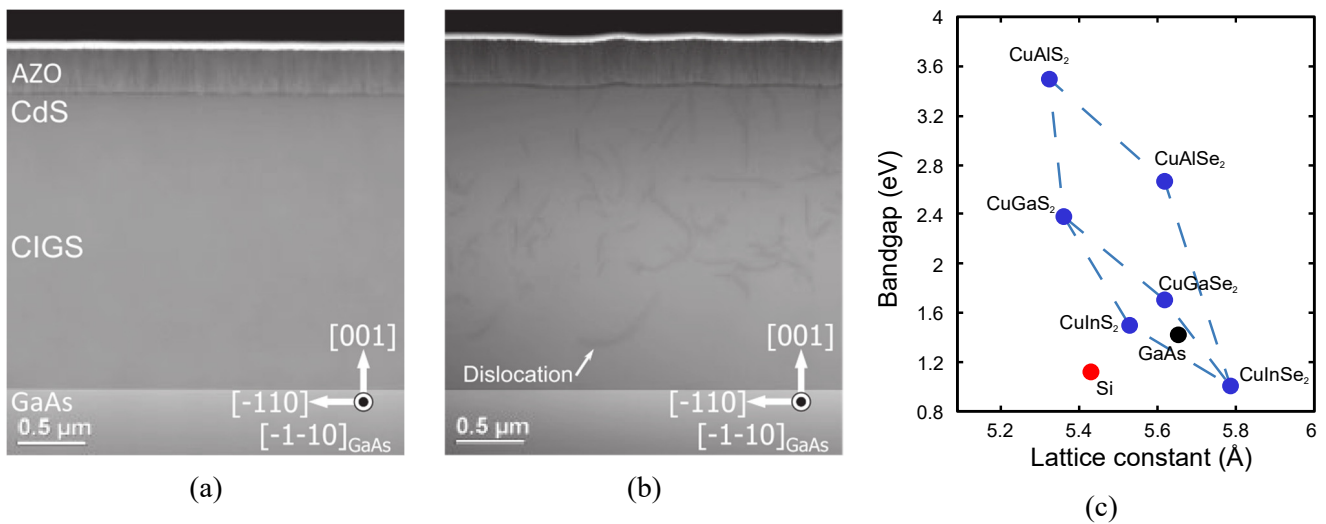

**Figure 12** – Bright field STEM images of epitaxial CIGS on a single crystalline GaAs substrate using (a) a lattice matched Ga/(Ga+In) = 0.6 composition, and (b) a top-bottom Ga/(Ga+In) grading from 0.7-0.4. The dislocations appear in (b) as a result of a lattice mismatch during growth. Both figures reproduced from [94] with permission, copyrigtht © 2018 The Japan Society of Applied Physics; (c) Bandgap versus lattice constant of selected members of the CIGS family of alloys.

In the case of CZTS, one of the most interesting aspects is that it is already nearly lattice matched with Si, with a mismatch of less than 0.1% for the sulfide $Cu_2ZnSnS_4$ [99]. This property should make the epitaxial growth of CZTS on Si quite interesting, and has been explored recently [100]. On the other hand, the bandgap of CZTS can be tuned in the ideal 1.6-1.8 eV range by cationic substitution, which is currently one of the most explored fields of kesterite research [101]. Nevertheless, this remains still a relatively unexplored field, mostly due to the lack of progress in the solar cell efficiencies of these high bandgap chalcogenide materials.

A fundamental limitation of this upright metamorphic growth is that the lattice mismatch is introduced on the very first layers being grown, and thus the resulting threading dislocations propagate to all the top cells. It has been estimated that the buffer layer needs to keep the threading dislocation density (TDD) below $10^6$ cm$^{-2}$ in order not to compromise the top cells [26,90]. This is only achievable for a restricted number of materials and configurations. For upright metamorphic III-Vs on Si, even though strategies to achieve a TDD down to $1\times10^6$ cm$^{-2}$ have been discovered over three decades ago [102], achieving this value in devices is considered impractical or exceedingly difficult, and typical TDD values remain near $10^7$ cm$^{-2}$ [103]. Therefore, this remains a major engineering challenge in order to combine high-efficiency III-V concepts with low-cost Si bottom cells. Interestingly, it was discovered that a derivative of CdTe alloyed with Zn, CdZnTe (in fact a II-VI group alloy), can be epitaxially grown lattice mismatched on Si and tolerate a much higher concentration of dislocations, still exhibiting carrier lifetimes higher than GaInP or GaInAs. A metamorphic 2J CdZnTe/Si tandem cell with 21.5% efficiency [104,105] has been demonstrated using this approach, by using molecular beam epitaxy (MBE) to grow CdZnTe. However, the high costs of MBE limit the prospects, and there has been relatively little work in this field in recent years. Moreover, despite the recent progress in fabricating Si-based tandems using this upright metamorphic growth, another fundamental obstacle is that, in all cases (III-V, CIGS, CZTS, CdZnTe), the growth of the top cells proceeds at high temperatures, which causes a plethora of integration problems:

- For instance in III-V deposition by MBE, the deposition temperature typically lies between 500-700 °C [93,106,107], which has the following implications:
  - Since Si and GaAs have different thermal expansion coefficients ($\alpha_{Si}$ = 2.6 × 10$^{-6}$ K$^{-1}$ versus $\alpha_{GaAs}$ = 5.7 × 10$^{-6}$ K$^{-1}$), thermally-induced stress is likely to occur during the upright growth from Si [106]
  - A diffusion of dopants from the tunnel junctions can occur, leading to loss of tunneling functionality. This requires either the use of diffusion barriers or lowering the temperature during III-V growth [93]
  - The high temperature limits the choice of Si bottom cell. For instance, the highest efficiency Si concept based on HIT structures cannot be used in this configuration because the amorphous Si passivating layers start significantly degrading after 200 °C, leading to depassivation of the heterointerfaces and increase in surface recombination velocity [108]
- CIGS and CZTS also require at least one high temperature step in their synthesis (>500 °C), which besides limiting the choice of Si bottom cell can also cause contamination of the Si bottom cell by diffusion of metallic elements, in particular Cu. Therefore, a diffusion barrier is likely needed, limiting the prospects of epitaxial growth. This has been a recent topic of research [109,110].

*3.2.4 Inverted metamorphic tandem cells*

To circumvent the lattice mismatch limitation, one very interesting strategy consists in inverting the metamorphic growth direction, that is, growing the first subcells lattice-matched, and only using metamorphic growth for the final subcell. This strategy is referred to as inverted metamorphic (IMM) growth. That way, the dislocation defects are confined to the top cell, because the dislocations follow the strain gradient [65]. Due to the limited choice of single crystalline substrates for lattice-matched epitaxial growth, this strategy also involves a substrate transfer process, as illustrated in **Figure 13**. The most successful example is growing a triple junction based on III-V top cells first on a GaAs substrate, and then using a buffer layer for metamorphic growth of the final junction (the bottom subcell). Then, the back contact is deposited (on the top layer), and a handling substrate is applied for mechanical support, using epoxy for adhesion. This handling substrate can either be permanent or temporary. Then, the GaAs substrate is removed using a lift-off or etching technique, the structure is flipped and the device is finalized with the front contact. Despite the higher complexity and steps involved in the transfer process, this method has two fundamental advantages. First, the bottom subcell (last subcell grown) can be grown epitaxially instead of by creating a diffused junction on a single crystal, as was the case on the Ge single crystal bottom cell of the classical InGaP/GaAS/Ge tandem mentioned above. This allows higher flexibility on the choice of the bottom cell, and allowed the transition away from single crystal Ge bottom cells, which led to a significant improvement in the efficiency records for triple and higher junctions. These types of 3J tandem structures have achieved 37.9% AM1.5G efficiency and 44.4% efficiency under concentrated light, and 4J, 5J and 6J configurations of this kind have further achieved 39.2% AM1.5G and 47.1% concentrated, making it the most efficient solar cell technology currently known to date [15,111].

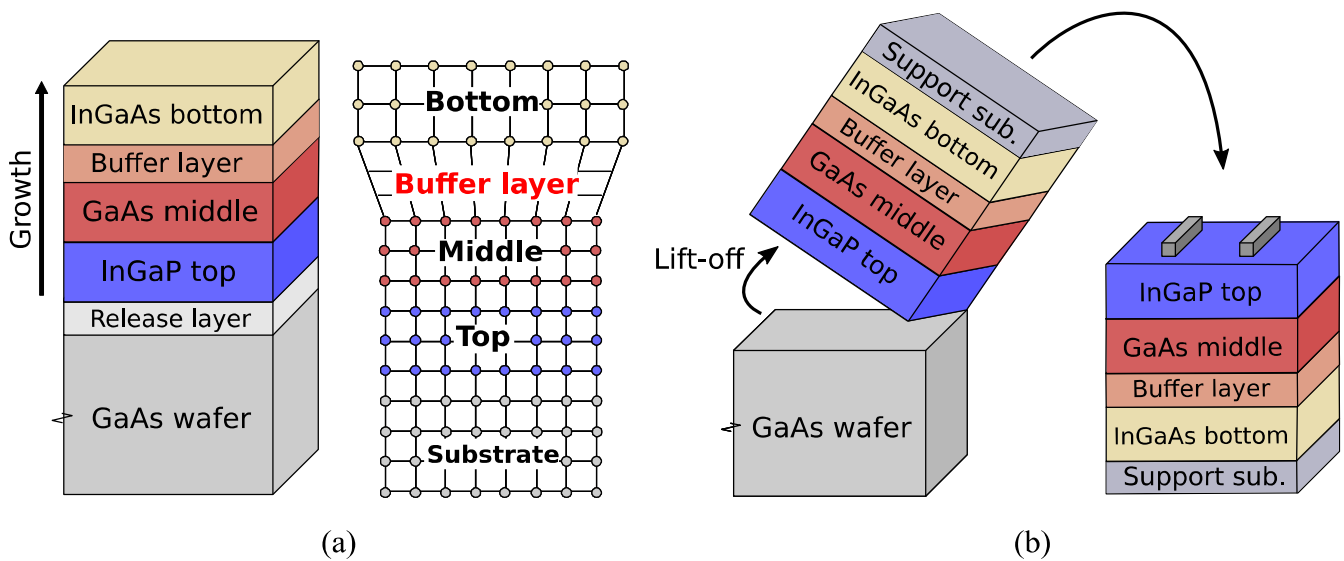

**Figure 13** – Processing sequence in the most successful inverted metamorphic (IMM) multijunction solar cells. First (a), the top subcells are grown lattice matched on a single crystal GaAs substrate, and the bottom cell is grown metamorphically. Then (b), the GaAs substrate is removed, and the thin-film layers are transferred to a supporting substrate, where the device is finished. Figure inspired by the work of Takamoto et al. [112].

Second, the removal of the growth substrate allows changing the device into a flexible and lightweight handling substrate, which is how the reduction from 2.6 kg/m$^2$ to 0.6 kg/m$^2$, mentioned in the beginning of this section, can be achieved to provide extremely high specific powers >3.6 W/g, crucial for space applications [113]. The integration of this IMM technology for space applications has been developed in the last decade [112], and in September 2013 this technology was tested for the first time on a satellite orbital test for 3.5 years. The results have been published recently, and confirmed its excellent radiation hardness and stability [113]. Note that due to its superior specific power metrics, these devices can also accommodate thicker cover glasses to prevent radiation damage without exceeding the tolerable weights of solar arrays [113]. For that reason, this technology is promising for future-generation space missions. The main disadvantages are that the higher process complexity involved in the substrate transfer and the requirement of several µm thick buffer structures impact the economics of this technology, the latter also applying to upright metamorphic structures. Nevertheless, the costs can be reduced by increasing the number of times the initial substrate can be reused [65].

*3.2.5 Wafer-bonded tandem cells*

Another variation of the lattice-matched and metamorphic approaches described above is wafer bonding. In the wafer bonding approach, two separate stacks are grown independently onto two substrates, and are brought together into a single monolithic device by bonding them into the other stack. A key difference here compared to the other techniques is that since the two stacks consist of rigid single crystalline wafers, there is always at least one rigid substrate for handling, meaning that there is no need to apply a handling substrate with epoxy [114]. However, this additional handling substrate is still required if the layer to be bonded is an intermediate (buried) layer instead of an exposed layer [115]. Besides this possibility, another fundamental advantage of the bonded wafer approach is that it relaxes the need for metamorphic buffer layers, as two different single crystalline substrates can be used for lattice-matched epitaxial growth. The bonding surfaces are treated either by chemical mechanical polishing (CMP) [114] or chemically activated by ion or atom bombardment surface treatments [107,115]. This way, the two bonded surfaces form strong covalent bonds with a strength similar to that of the atomic bonds in the other layers [114], and lead to transparent, durable and low-resistance interfaces [65,114]. The two wafers are mechanically pressed, forming an almost uniform bonding area as shown in **Figure 14 (a)**. Then, the upper substrate is removed, so far most successfully by complete chemical etching of the substrate [106,107,112,114,116], or by substrate lift-off after chemically etching a sacrificial layer (epitaxial lift-off or ELO) [117,118]. Other possibilities include ion implantation, laser- and stress-induced lift-off, which are common in other fields but less explored here [65]. Together with the IMM approach, this technology has enabled the highest solar cell efficiencies to date, with 38.8% AM1.5G efficiency for a 5J tandem and 46.0% for a 4J under concentrated light [111].

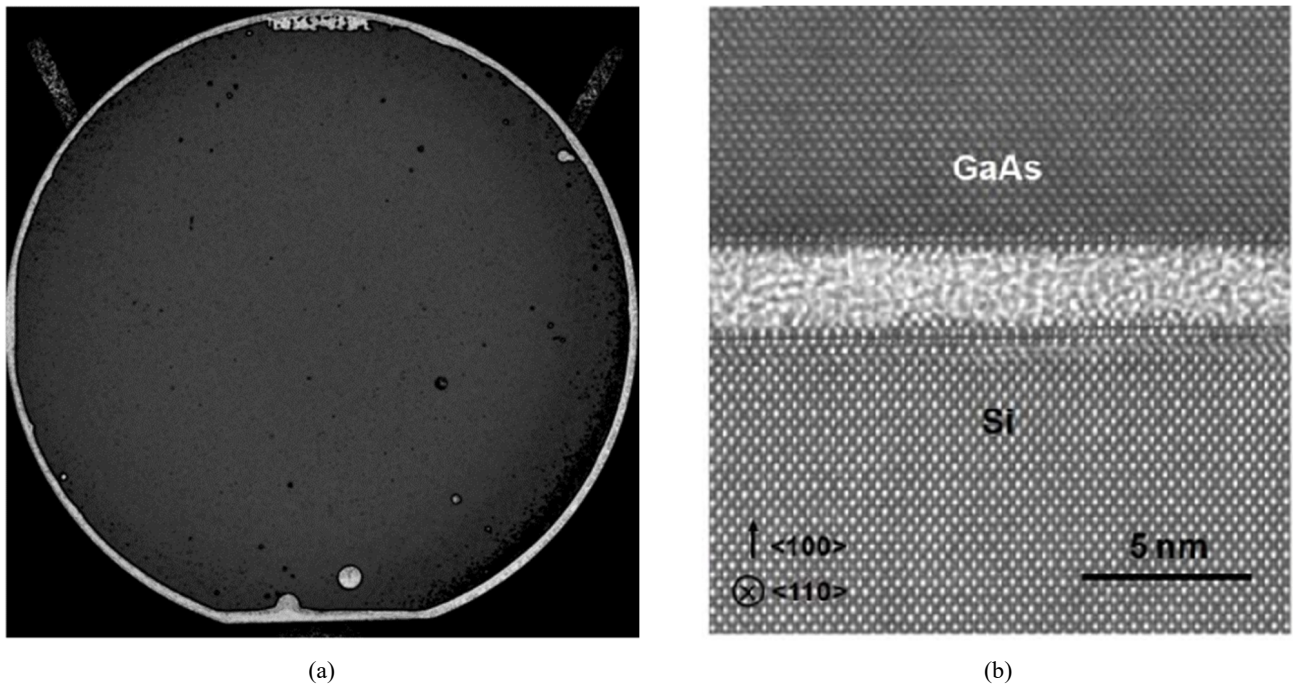

(a)  (b)

**Figure 14** – (a) Scanning acoustic microscope image of a bonding between a GaAs and a Ge wafer, with 4 inches in diameter. The wafers are completely bonded except for some small bubbles shown in black. Figure reproduced with permission from Niemeyer et al. [114]; (b) Transmission electron microscope (TEM) cross-section image of the amorphous layer at the bonding interface between GaAs and Si. Reproduced with permission from Tanabe, Watanabe and Arakawa [119].

From the point of view of combining the highly efficient III-V compounds with Si, wafer bonding is perhaps one of the most promising paths, because the 4.1% lattice mismatch between Si and GaAs can be directly bypassed by bonding a III-V stack onto a Si device wafer. When the surfaces are appropriately treated, a stable bonding between GaAs and Si can be achieved, forming a very thin interfacial amorphous layer, as shown in **Figure 14 (b)**. Therefore, the two technologies can be combined without the need for a buffer layer and without the occurrence of detrimental dislocation defects. Furthermore, the high-temperature growth is no longer as limiting, since the two stacks are developed separately before bonding. In any case, a tunnel diode between Si and the first III-V subcell is still required, and is usually done on the III-V structures. With this wafer bonding strategy, Si-based 2J devices have reached 30.2% AM 1.5G efficiency [116], 30.0% under concentrated light [106], and 3J III-V bonded on Si have achieved 34.5% under AM 1.5G, the latter achieved very recently at the end of 2020 [92]. This efficiency improvement represents a 15-30% relative improvement over the best single-junction Si cells. Given the direct impact of efficiency on costs mentioned in the previous section, this makes wafer-bonding one of the best candidate strategies for delivering low PV costs. Perhaps the biggest challenge for reducing the cost of this approach is that, so far, the best devices were achieved by chemically etching the whole GaAs substrate after bonding, which therefore cannot be reused to grow new III-V layers. Unless the price of GaAs wafers significantly decreases, this destruction of GaAs wafers will imply high costs. Instead, a lift-off process such as the ELO could provide reusability and better cost metrics, but so far there have been no public demonstrations of more than 100 reuses [16]. Moreover, a surface conditioning treatment involving a very expensive polishing step (CMP) is often required to achieve that reusability.

For the sake of completion, we mention that other fields of research exist in III-V multijunction devices, which also try to overcome the limitation of the classical 3J InGaP/InGaAs/Ge cell. One of them consists in introducing quantum wells (QWs) in the middle cell in order to tune the absorption characteristics and provide current matching. Another method consists in expanding the classical 3J into a 4J and 5J configuration, still with Ge as bottom cell, in order to achieve a better current matching between all the subcells. Due to the remarkable progress in the metamorphic and wafer bonding methods, the interest in these fields has declined in the last years [65]. However, using this QW incorporation strategy, a new 2J one-sun efficiency record of 32.9% has been recently achieved for a GaInP/GaAs cell grown using purely lattice-matching strategies (non-metamorphic) [120], which might give some new research prospects. A third method is based on nanowire solar cell concepts, where III-V nanowires are used as a way to relax the lattice mismatch by straining the wires. This method is relatively new and still in the development phase [65].

*3.2.6 Mechanically stacked tandem cells*

Finally, another method that has been investigated to integrate III-Vs on Si consists in mechanically stacking the III-V stack on a Si bottom cell, and achieving an electrical connection by wiring instead of wafer bonding. Three examples are shown in **Figure 15 (a)-(c)**. This approach has several fundamental advantages. First, the direct wiring eliminates either the need for metamorphic buffer structures from Si to the III-V structures, or the need for wafer bonding and the associated surface treatments necessary, which always have to be performed in a cleanroom due to the requirement of very clean surfaces. Second, it eliminates the need for a tunnel junction in that same interface between Si and the III-Vs. Third, since the top and bottom cells are joined by a glass (usually using an insulating polymer adhesive), this approach is essentially modular in the sense that any kind of top cell can be fabricated on that glass. Therefore, no major specific changes are required to make the subcells compatible for tandem integration, and high-temperature synthesis methods are possible. This makes this configuration the most flexible of all the configurations mentioned above, in terms of design choices for both the bottom Si and the top cells. Because of this higher simplicity, this configuration has achieved the best multijunction efficiencies of III-V on Si, with 32.8% for a 2J and 35.9% for 3J under AM 1.5G conditions [16]. Despite its inherent functional advantages, the mechanical stacking method also needs the same substrate transfer procedure as the IMM method, and therefore wafer reusability is still a concern. On the other hand, the cell interconnection is now not monolithic and therefore requires additional wiring. This includes additional grid electrodes in the Si/III-V interface, which reduce the bottom cell transparency, and also wiring between the subcells, which makes the upscaling to module integration more complicated and costly. To the best of the author's knowledge, this upscaling feasibility has not been specifically addressed yet for this technology.

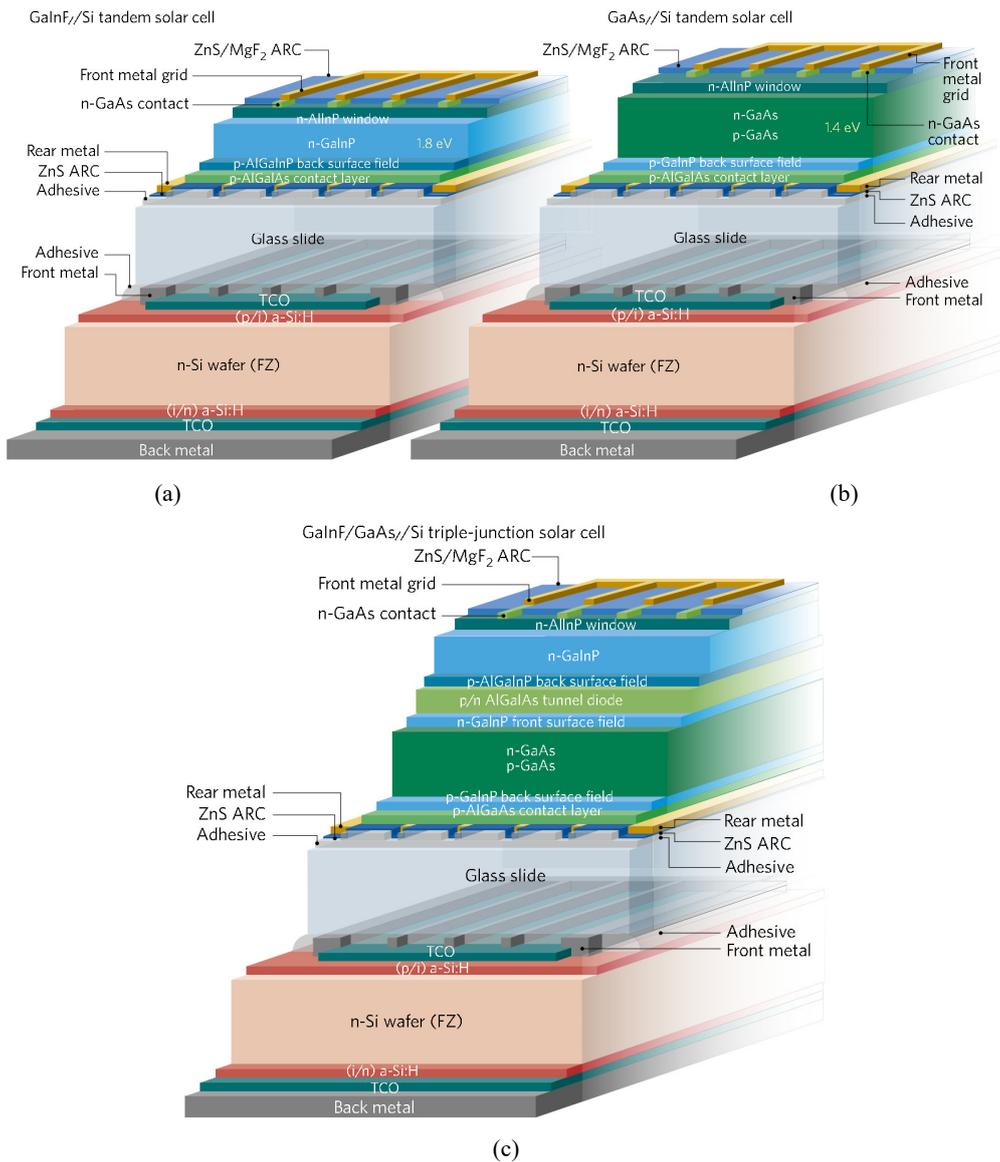

**Figure 15** – (a) 2J mechanically stacked tandem cell based on GaInP on Si; (b) 2J mechanically stacked tandem cell based on GaAs on Si; (c) 3J mechanically stacked tandem cell based on GaInP/GaAs//Si. Figures from [16].

Several cost models have been developed to estimate the potential of achieving low-cost tandem devices for transitioning the terrestrial photovoltaic industry into a multijunction technology. Due to the extremely low costs of the Si technology, it is believed that the lowest possible costs will be achieved using Si-based devices rather than devices based exclusively on other technologies, such as III-V subcells. This can be understood from our analysis above. In a III-V on Si device, the number of III-V wafers required would be at least one order of magnitude lower than the number of Si wafers required, due to the wafer reusability for III-V epitaxial growth. This value could potentially be two orders of magnitude if reusability is extended to above 100 times. In contrast, a III-V tandem device using a III-V wafer as active cell would always require at least one III-V wafer per device. Considering the very high costs of III-V wafers, this would mean that the costs would be at least one order of magnitude higher just on wafer costs alone. In the IMM tandems, cheaper substrates can be used and the III-V substrate can be reused, but the thick metamorphic grading layers add to the overall costs. Based on this, it can be predicted that the lowest possible costs (and the highest efficiency-to-cost ratio) are more likely to be achieved when III-Vs are combined with Si. Such a cost forecast is shown in **Figure 16**. Some of the most important cost-reduction aspects described by the model are the wafer reusability without the need for CMP, as mentioned, and an overall reduction in the substrate costs and the cost of the III-V deposition methods. The model also assumes that an efficiency up to 35% can be achieved long term. The long-term cost prediction of $0.66 is only 3 times higher than the 2019 cost of PV, and would correspond approximately to the cost of PV in 2015. With further improvements, we could therefore see this technology implemented in terrestrial applications, in particular in conditions where a higher power density or limited area available would justify the cost premium, as well as in hot environments, due to their superior temperature coefficients. In this case, coupling the tandem device with concentrating optical systems to further increase the energy yield could also be justified [29]. In particular, this approach could also lead to cost-efficient modules by reducing the active areas required, as discussed in the next section.

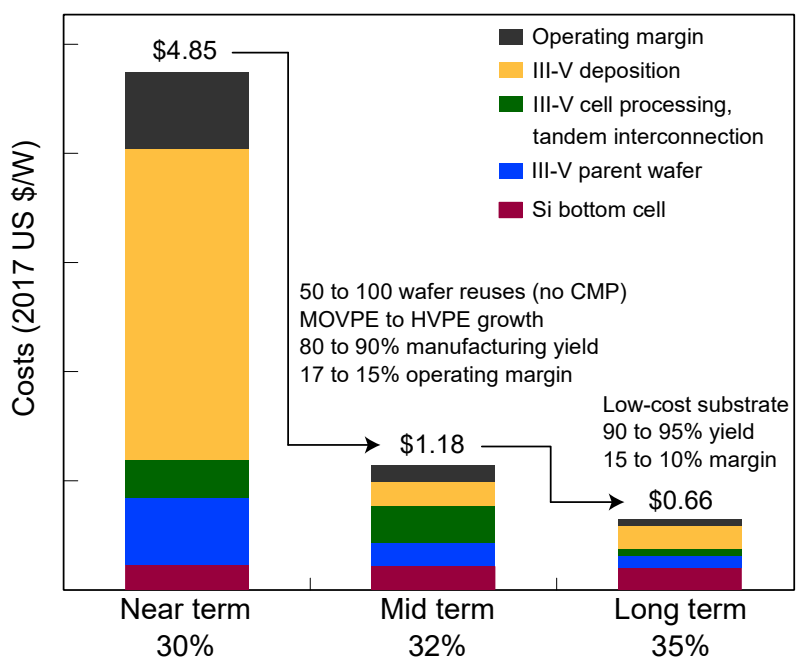

**Figure 16** – Cost forecasts for multijunction solar cells based on III-Vs on Si. Figure reproduced from [16].

*3.2.7 Concentration photovoltaics with tandem cells*

Of all the technologies described in this work, concentrating sunlight on a solar cell by optical means is, by far, the method that achieves the highest energy in a given active cell area. Concentration photovoltaics (CPV) surpasses the original SQ limit by reducing the solid angle mismatch between the sun and the cell, and therefore, in ideal conditions, the actual efficiency also increases along with the increase in energy density. At the maximum theoretical concentration ratio of 46200 (i.e, 46200 times the non-concentrated solar irradiance), the maximum efficiency reaches 45% for a 1J device with a bandgap near 1.0-1.2 eV [81]. For multijunction devices, detailed balance efficiencies well above 50% are possible even at a realistic concentrations of 100-1000 [121]. Most importantly, this efficiency refers to a larger incident irradiance, meaning that the actual energy yield per area is orders of magnitude higher. For example, a non-concentrated 1J device with 25% efficiency produces 250 W/m$^2$. In comparison, the highest efficiency concentration cell reported, a 6J cell at 143× concentration with 47.1% efficiency [15], achieved a power density of 67464 W/m$^2$, nearly a factor of 270 higher. In practice, this means that the same power output can be achieved with similar modules but much smaller cells. This trade-off could compensate the overall costs and allow the use of very expensive III-V cells which, as mentioned above, are currently still at least 1 order of magnitude more expensive than Si.

Ideally, $J_{sc}$ scales linearly with irradiance, and the open-circuit voltage ($V_{oc}$) scales logarithmically. However, the effect of series resistance ($R_s$) becomes a fundamental limitation in practical devices. Taking single-junction Si as an example, for a 500× concentration, a $J_{sc}$ above 20 A/cm$^2$ can be obtained [122]. At this level of current, typical metallization schemes used in conventional Si cells, with $R_s$ on the order of 0.1-1 Ωcm$^2$ [34] would lead to a voltage drop higher than 1 V due to $R_s$ alone, leading to a complete loss in power output. For Si at 500×, to operate close to typical open-circuit voltages of 800 mV, the $R_s$ should be kept below 0.001 Ωcm$^2$, corresponding to less than 3% relative loss in power [123]. So far, only $R_s$ values down to 0.002 Ωcm$^2$ have been demonstrated for Si using specific point-contact IBC metallization structures [122]. For multijunction devices, the situation is more favorable, because the total current in the device is lower, and the voltage is higher instead. In a 2J tandem, the total $J_{sc}$ is approximately halved compared to a 1J device. Similarly, in a 6J device, the one-sun $J_{sc}$ is reduced to around 9 mA/cm$^2$. As the power loss due to series resistance scales with the square of the current density, this relaxes the tolerable series resistance by a factor of roughly 20. And indeed, experimental III-V tandem devices have achieved $R_s$ values down to 0.015 Ωcm$^2$ [124], making them the most promising candidates for CPV applications. The higher the series resistance, the faster the efficiency drops as a function of concentration. On the other hand, high concentrations are needed to compensate for the higher cell costs, which is the fundamental trade-off limiting the viability of this technology. **Figure 17** (a) and (b) illustrate simulations of this trade-off for the efficiency as a function of series resistance in a 3J tandem. In practice, due to this effect, the concentration factor has been limited to below 1000× in current devices. Still, 3J III-V tandem devices have achieved up to 40.9% efficiency at 1093× [125], which represents a remarkable power density of 402333 W/m$^2$, more than 3 orders of magnitude higher than standard 1J technologies.

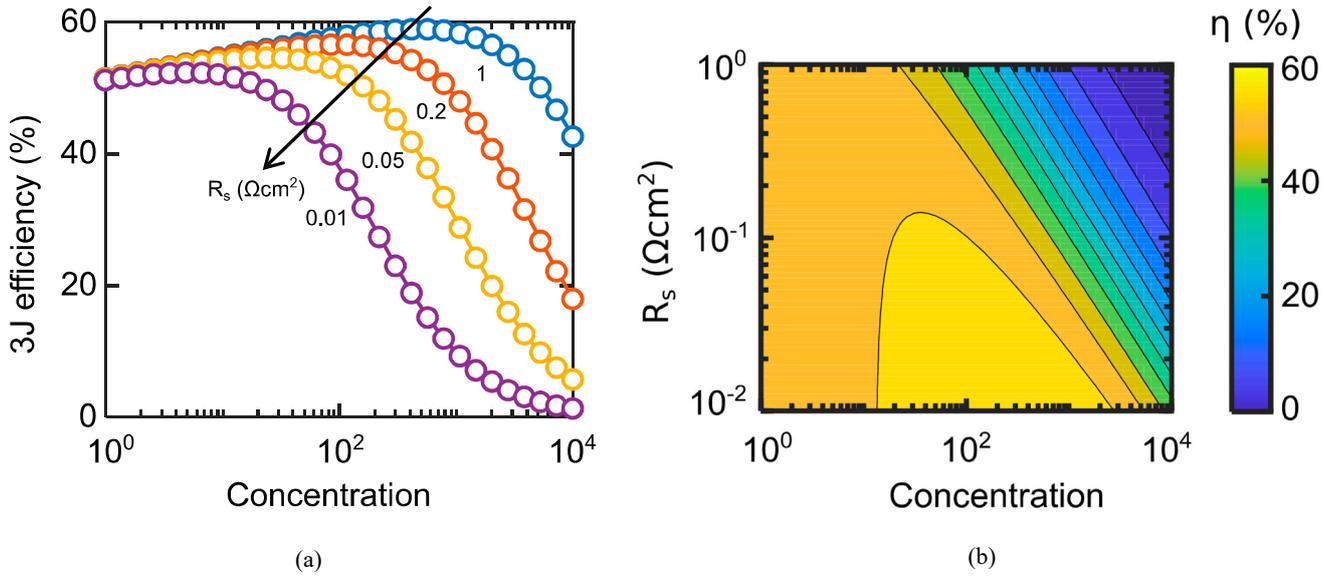

**Figure 17** – (a) Detailed-balance efficiency limit of a triple-junction tandem as a function of light concentration for different values of the series resistance; (b) corresponding efficiency color map for the same system. Both figures adapted from [126] with permission, copyright © 2019 National Academy of Sciences.

The reduction in cell dimensions can then allow a restructuring of the module to accommodate the optical focusing elements and a temperature sink. Due to the high light concentration and power dissipation due to series resistance, high-capacity heat sinks are required to cool the cells, and in practice the operating temperature is always above 60 °C [125]. These high temperatures reduce the photovoltaic performance and can lead to system damage, which further hinders high concentration ratios. On the other hand, as we have discussed, tandem devices have the advantage of a better temperature sensitivity compared to single-junction devices. In fact, this advantage is even more prominent in concentration applications, as concentration reduces the Boltzmann losses, which have the highest temperature coefficient of all detailed balance loss mechanisms [75]. This improvement in temperature sensitivity has been confirmed experimentally [127], and means that high-efficiency tandem devices exhibit excellent temperature stability under concentrated light.

There have been considerable efforts in commercializing CPV systems, and over 360 MW of installed capacity was documented worldwide as of 2016 [128]. Due to the advantages of III-V multijunction devices, concentration systems based on 1J Si have been discontinued or only used in low concentrations (<100×) since 2005 [122], and over 90% of the installed capacity is based on high-concentration III-V tandem systems [128]. Due to the successful cost reductions of the current 1J market technologies, commercial developments in the CPV sector have decreased. However, considering the recent progress in III-V tandem devices, there are promising prospects of market expansion. In particular, it has been estimated that at 30%/1000× levels, the III-V raw material supply levels could allow more than 1GW/yr of production [128]. Moreover, CPV systems can be combined with utility-scale concentration solar thermal systems, which also make use of light concentration. On the other hand, the abovementioned power density figures are given assuming the AM1.5D standard, where only direct radiation with 900 W/m$^2$ is considered. Diffuse or scattered light cannot be concentrated, which means that CPV systems severely underperform under non-ideal spectral conditions. For that reason, hybrid CPV systems have been suggested, where a CPV system can be combined with a standard Si cell to use non-direct light [129]. Such a III-V/Si hybrid system was recently introduced as the first hybrid module in the PIP efficiency tables with a combined efficiency of 34.2% [92].

*3.2.8 Emerging tandem cells and alternative subcell interconnections*

Recently, the advent of highly efficient metal halide perovskite solar cells based on methylammonium lead halides has brought new approaches to fabricating Si-based tandem cells. Due to the remarkable defect tolerance of these perovskite compounds, they can be deposited as polycrystalline thin films on Si without the need for epitaxial or lattice-matched growth, and still exhibit carrier lifetimes on the order of microseconds [33]. This high material quality can be obtained using a wide range of cheap and simple deposition methods. Moreover, the material crystallizes at around 100 °C, meaning that it can be directly deposited on a fully developed SHJ/HIT Si solar cell without degrading it [33]. Because of the excellent photovoltaic properties of the perovskite and its very low processing temperature, this allows for new tandem interconnection configurations. Chang et al. [130] recently developed a bottom-up cost analysis for some notable monolithic perovskite/Si configurations reported in literature. In **Figure 18**, we expand on this study, with a particular focus on the subcell interconnection approach. In general three different interconnection strategies have been proposed that yield promising results. The first one consists in having the perovskite and Si sharing a common carrier membrane – either a hole transport layer (HTL) or an electron transport layer (ETL). This can be achieved by either SnO$_2$, illustrated in cases **(a)** and **(b)**, or by TiO$_2$, case **(c)**, which simultaneously act as ETL for the perovskite and as recombination junction at the interface with Si. In particular, a low resistance TiO$_2$/p+ Si interface can be achieved by tuning the band alignment and respective doping densities [131], and therefore this is a case-specific recombination junction suitable for PERC Si bottom cells. The second configuration consists in using a transparent conductive oxide (TCO) as interface recombination layer, separating the top cell structures from the bottom cell. The most commonly used TCO is indium tin oxide (ITO), as shown in cases **(d)-(g)** and **(i)**, although other TCOs can be used, in particular fluorine-doped tin oxide (FTO), which has the benefit of higher thermal resilience. Due to its generally favorable ohmic contact to heavily doped Si structures, TCOs can be used with different Si bottom cell configurations besides the PERC, such as the SHJ/HIT structure, based on hydrogenated amorphous Si (a-Si:H) layers, as shown in cases **(e), (f), (h)** and **(i)**. On the other hand, perovskite devices are routinely developed in superstrate configuration on ITO-coated glass substrates. As a result, the TCO configuration is the most commonly used so far in perovskite/Si tandems. The third possibility is using a tunnel junction, as described above, but this time fabricated on the Si side by means of a modified SHJ structure, as illustrated in case **(h)**. Another notable aspect of this structure is that the top cell fabrication could be achieved for the first time on a double-side textured Si bottom cell, in contrast to other configurations where a polished front side is required for the spin-coating of the perovskite top cell layers [132]. With these different integration strategies, enabled by the relatively soft and simple fabrication processes of perovskite top cells, perovskite/Si tandem cells have now surpassed the efficiency of single-junction Si, with a recent new record of 29.52% being certified at the end of 2020 [21]. The corresponding tandem configuration for this record cell is illustrated in case **(i)**. This technology demonstrated, for the first time, that it is possible to achieve beyond-Si efficiencies with relatively simple device architectures and polycrystalline growth methods, without the need for lattice matching or even epitaxial growth. These architectures have been applied to perovskite/perovskite [133] and perovskite/CIGS tandems [74], in both cases reaching already certified efficiencies of 24.2% [111].

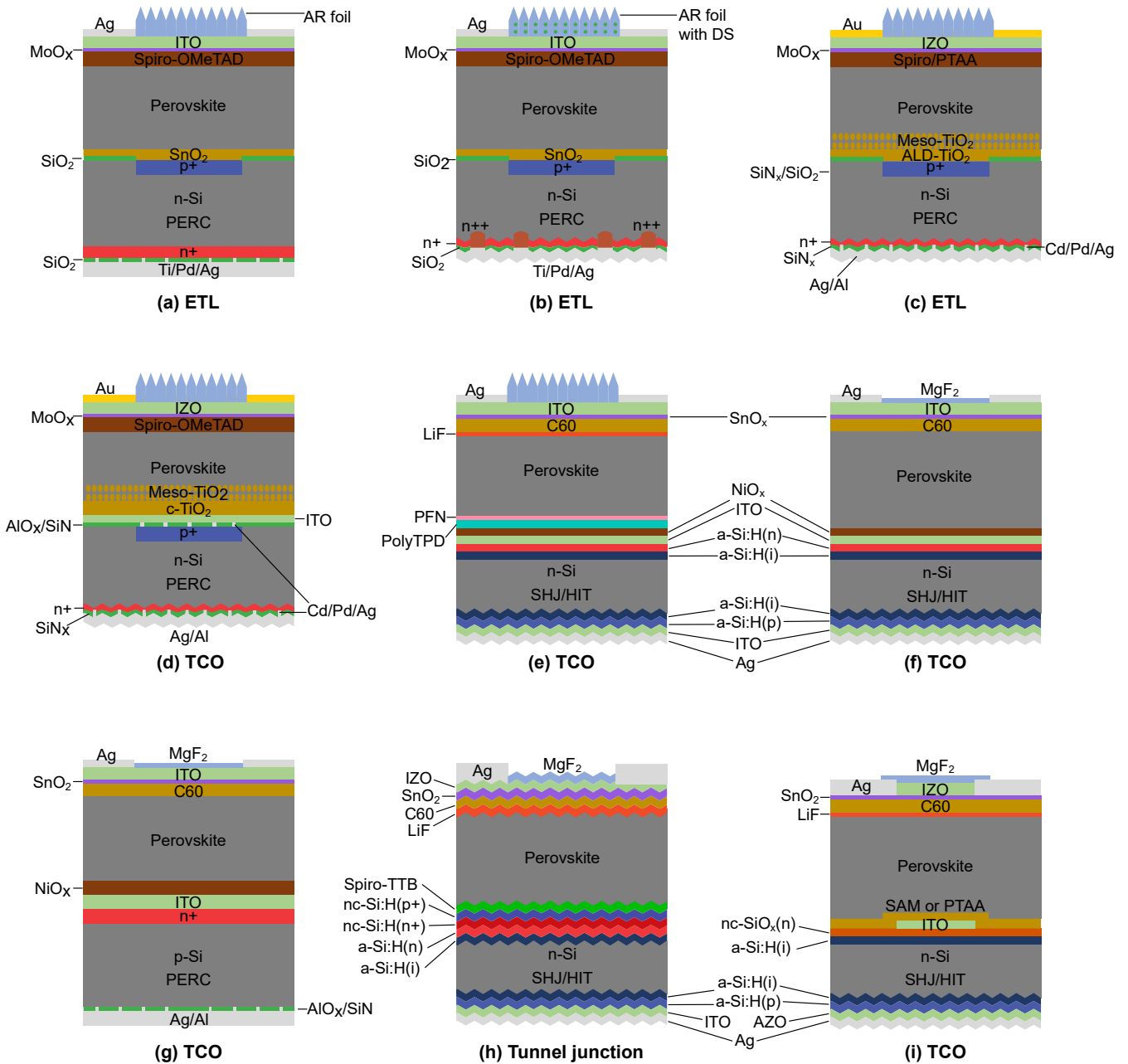

**Figure 18** – Notable examples of different monolithic perovskite/Si tandem configurations organized by the type of recombination junction. The layers are not to scale. Expanded based on the work from Chang et al. [130] with permission, copyright © 2020 John Wiley & Sons.

Due to this remarkable success, a number of startups have sprung aiming to upscale the development of perovskite/Si tandem solar cells, namely Oxford PV, Swift Solar, Tandem PV and Saule Technologies. Similarly, a lot of interest has been raised in the PV scientific community, with over 3600 and 3800 publications in 2019 and 2020, respectively, and a total number of 17000 cumulative publications so far in December 2020.[1] Currently, this is the most promising way to achieve the highest Si-based tandem efficiencies at the lowest possible costs. In their bottom-up cost model, Chang et al. suggest that a hypothetical simplified tandem structure, shown in case **(g)**, with a minimum of 25% efficiency, could yield a lower LCOE than single Si cells [130]. Moreover, it is expectable that other perovskite/Si tandem configurations will appear in the future, for example based on other high-efficiency Si bottom cells such as the POLO/TOPCon. However, the current limitations of perovskite solar cells are that the highest efficiency variants use the toxic element Pb and have stability problems, suffering from multiple instability or degradation sources such as humidity, oxygen, temperature, electrical bias and light, some of which remain poorly understood. Recently, a consensus statement from multiple authors has been issued [134] seeking to provide guidelines for addressing these stability issues and unify the measurement protocols. As reviewed there, although several perovskite devices have surpassed internationally-established standard stability tests, the longest stability under operation of these perovskite compounds remains on the order of 2000 hours, or approximately 0.23 years, still very far from the 30+ years achieved by crystalline Si.

---

[1] Calculated by searching for the simultaneous occurrence of the terms "Perovskite" and "solar cell" in the Scopus database.

Other possible interconnection configurations have been proposed in organic-based tandem devices. Ag or Au, either as nanoparticles or as nm-thick layers have been proposed as recombination junctions, due to their reasonable transparency and protection of the bottom organic structures [31]. Similarly, other conductive matrices have been suggested, such as a nanoparticle mix of ZnO and poly(3,4-ethylenedioxythiophene) (PEDOT) [135]. In Si-based tandems, thin TiN-based conductive layers have also been proposed as alternative to TCOs, with the double effect of recombination layer and diffusion barrier protecting the bottom Si cell during top cell growth [109,110]. In all cases, this intermediate connection should satisfy the condition that the Fermi level equalization between subcells is achieved without voltage losses, i.e. with low resistance and without carrier extraction barriers or interface recombination. As we have described, some configurations are case-specific, and the softer the fabrication processes of the top cell (i.e. low fabrication temperature), the wider the choice of interconnection configurations.

Given these remarkable recent advances described in this section, the current industry forecasts are that Si-based tandem modules might be introduced in 2024 [4] . Based on our discussion, the top cell technologies most likely to be a part of that Si-based tandem are III-Vs or Perovskites. Besides these, there is currently no known semiconductor with a bandgap around the ideal 1.6-1.8 eV range with satisfactory PV properties to achieve high-efficiency top cells. Therefore, one of the most important fundamental and theoretical research questions is discovering possible new candidates for such a material. This process could be accelerated in the future by using computational screening algorithms, some of which have already been suggested [136–138]. Interestingly, this not a problem limited to photovoltaics, as it is also a topic of major interest in the field of photoelectrochemistry, where solar photons can be combined with catalysis for a sustainable production of fuels and chemicals [139]. Here, tandem devices are equally desirable, as they extract more energy from the sun, and low-cost crystalline Si photovoltaic technologies could also play an important role if they can be properly adapted to electrochemistry.

*3.3 From cell to module operation: wiring and spectral variations*

It was implicit in our discussion above that the different multijunction configurations presented will result in different electrical configurations, namely in terms of the number of electrical terminals of the device. Due to their fabrication sequence, the upright metamorphic, inverted metamorphic and wafer bonding concepts used in III-V growth can naturally result in two-terminal (2T) devices. Moreover, the resulting devices are also considered monolithic, even in the cases where there is a transfer of substrate, because the final device is settled on just one single large-area substrate, such as a wafer or a flexible foil. Likewise, the above examples of perovskite growth directly on Si tunnel junctions, TCOs or other recombination layers also result in 2T monolithic devices. Therefore, all these concepts are cataloged as monolithic two-terminal devices, optically and electrically in series. Being electrically in series, the voltage output of each subcell is added, and the tandem current is the lowest current of each subcell. For that reason, 2T configurations have the constraint that the subcell currents need to match, a condition known as current matching, described above for the classical InGaP/InGaAs/Ge cell. However, for a mechanical stacking approach, there are four free terminals (4T), which can be reduced to a 2T if the top and bottom cells are connected in series, or otherwise operated in parallel. In general, we can divide all the possible configurations in the 6 cases illustrated in **Figure 19**. Of these, we have discussed the 2T monolithic, 2T mechanical stack and 4T mechanical stack in the previous sections. The 4T optical couple is a non-monolithic system where a dichroic filter is used to split the solar spectrum, matching the high wavelengths with the low bandgap absorbers, and the low wavelengths with the high bandgap absorbers. Notably, this system is a multijunction configuration but it is not a tandem device, because the subcells are not physically in tandem. Despite its obvious advantage of allowing independently built and operating devices, it is unclear whether this configuration can be scaled to large area modules due to the costs of the optical system, additional wiring, and the complicated logistics of arranging and integrating the optical system into a module. For this reason, we have left this configuration out of this review.

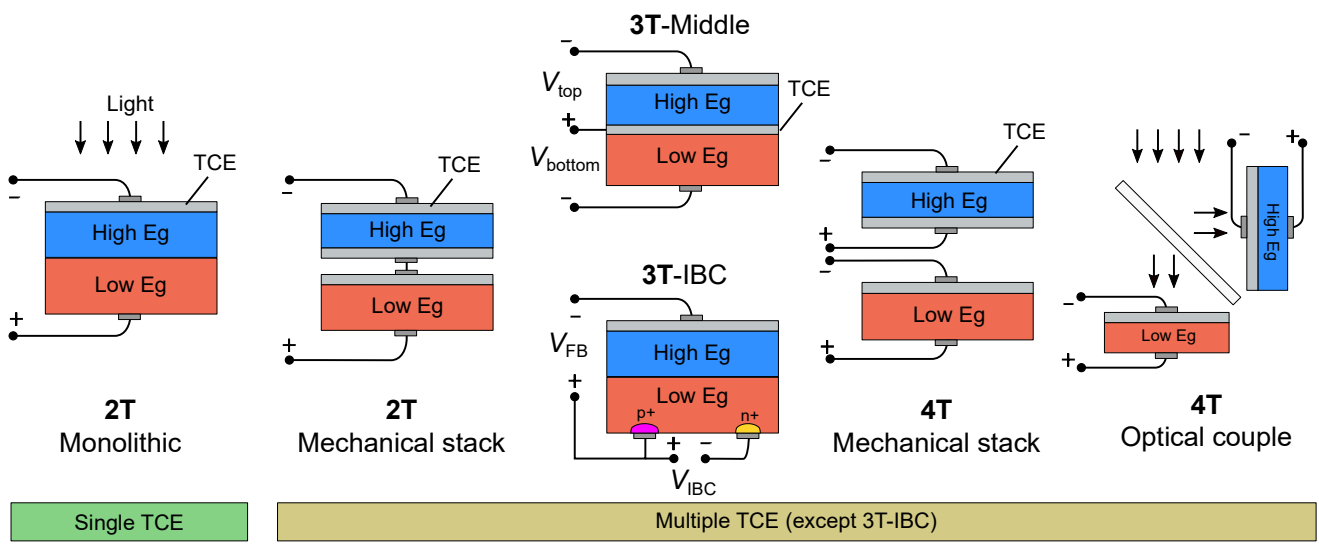

**Figure 19** – The catalog of different multijunction configurations, sorted by the number of electrical terminals and by the number of transparent conductive electrodes (TCEs). Note that despite the designations "High Eg" and "Low Eg", this catalog is also valid for a higher number of subcells.

The remaining notable cases are then three terminal (3T) configurations. One type of 3T device consists in introducing an additional electrode in the middle of the top and bottom cells, therefore labeled 3T-Middle. Because the middle electrode is common to both the top and bottom cells, they are no longer strictly in series. Instead, the cell has three interdependent terminals, like a transistor. If the top and bottom terminals are connected to different matching loads, this relaxes the current matching constraint. On the other hand, if the top and bottom terminals are connected together outside the cell, the top and bottom cells form a parallel connection. Another interesting aspect of the 3T-Middle configuration is that it allows a nearly independent measurement of each subcell. In particular, the external quantum efficiency (EQE) of both subcells can be accurately measured without the need for bias light [140]. For that reason, this configuration is quite attractive in a research phase when the performance of both subcells needs to be evaluated separately. For triple or more junctions, this can be achieved with intermediate electrodes between each subcell. Naturally, these electrodes need to be sufficiently conductive to avoid lateral transport resistance losses. However, a major drawback of this configuration is that the additional exposed electrode reduces the active area of the cell.

The other configuration can be obtained when the bottom cell has an interdigitated back contact (IBC) structure, where two electrodes with different polarities are fabricated on the backside of the bottom cell. This configuration has raised a lot of attention in the last 5 years due to its remarkably interesting properties, so it is analyzed in greater detail here. The equivalent electrical circuit of the 3T-Middle and the 3T-IBC configurations is essentially very similar, but the way the corresponding devices are fabricated can be quite different. In the 3T-IBC system, there is a circuit defined by the Front-Back (FB) connection between the front electrode of the top cell and the electrode of the bottom cell of opposite polarity, and an IBC circuit, defined by the interdigitated contacts of the bottom cell. It was discovered by Nagashima et al. in 2000 [141] that the additional degree of freedom provided by the bottom IBC extra circuit eliminates the need for current matching between the subcells, as with the 3T-Middle configuration. If the top and bottom cells are perfectly current matched, the FB circuit behaves as a real 2T device, and the IBC circuit does not output any power. If either of the subcells becomes current-limited, the excess current can be collected by the IBC circuit. This allows the tandem to always operate near its theoretical maximum output, regardless of temperature and spectral conditions, similarly to a real independent 4T operation. The 3T-IBC tandem is therefore significantly more resilient against spectral variations than a 2T configuration, as shown in **Figure 20 (a)**. In fact, spectral variations are one of the main fundamental drawbacks of 2T configurations for real-world operation. Any asymmetric changes in the solar spectrum, such as daily and seasonal changes in solar inclination (and therefore air mass), or atmospheric attenuation (e.g. clouds) will cause a mismatch in the subcell currents, thereby partially undoing the benefit of higher tandem efficiencies compared to 1J devices. There are still limited studies on the influence of spectral variations on the energy yield of tandem devices across different locations on Earth, but some studies [142,143] found that this mismatch can represent a 10-15% relative reduction in annual energy yield compared to a 4T configuration, based on device simulations using real weather data. On a 3T-IBC configuration, these mismatch variations can be compensated by the IBC circuit, as illustrated in **Figure 20 (b)**.

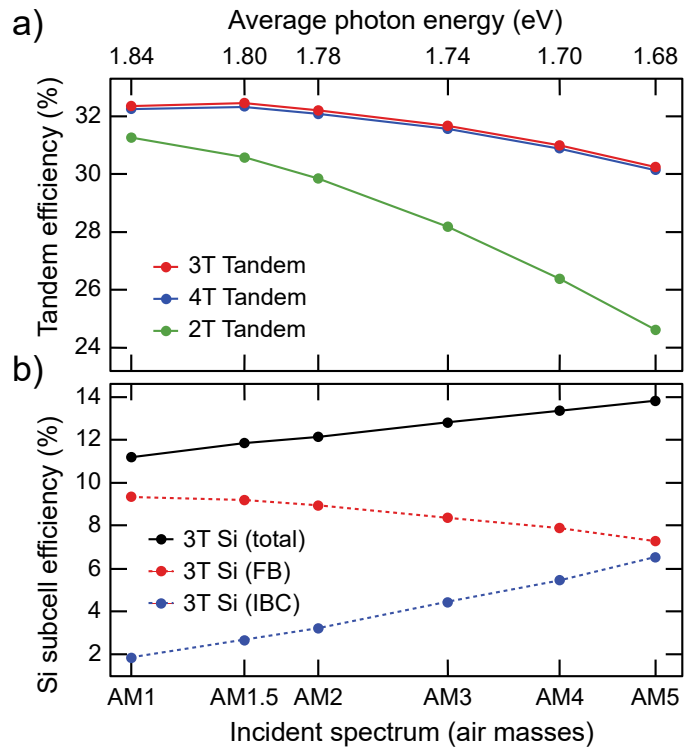

**Figure 20** – (a) total tandem efficiency simulations for a GaInP/Si tandem cell under different spectral conditions; (b) respective Si subcell efficiency divided by FB and IBC circuits. Figure adapted from [144] with permission, copyright © 2018 Royal Society of Chemistry.

In order for this nearly independent operation to be possible, the resistance for current injection between the top and bottom cells in the FB circuit needs to be minimized. This can either be achieved by a tunnel junction between the top and bottom cells, where the carriers can recombine, or by a carrier selective layer that allows the injection of one carrier type into the bottom cell, where it can recombine with its counterpart. The latter case is essentially the mode of operation of a bipolar junction transistor (BJT), where the font electrode is the emitter, the opposite IBC electrode is the base and the other IBC electrode acts as the BJT's collector [145]. These modes of operation are summarized in **Figure 21**. The different modes of operation for other variations of the 3T have also been reviewed by other works [144,146].

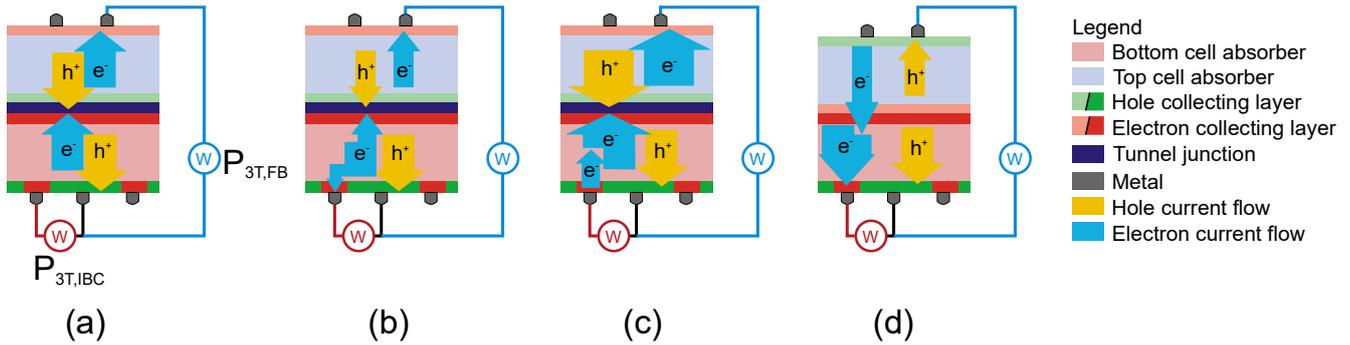

**Figure 21** – Modes of operation of a 3T-IBC tandem: (a) current matched subcells and standard 2T operation, with $P_{3T,IBC} = 0$; (b) Limiting top cell, with $P_{3T,IBC} > 0$; (c) Limiting bottom cell, with $P_{3T,IBC} < 0$; (d) Bipolar junction transistor mode, allowing current injection from the top cell into the bottom cell. Figure adapted from [145] with permission, copyright © 2019 John Wiley & Sons.

Therefore, the two subcircuits allow each subcell to operate at its maximum power point, and the resulting combined power output is essentially the same as a 4T device [144,145]. This means that the inverter associated with these solar cells needs to have two load matching circuits to operate each subcircuit at its maximum power point, instead of just one as in the standard case. In practice, since the tandem cells have to be wired in strings to obtain the standard voltage and current outputs of a commercial module, this means that the inverter will operate with two different substrings. This can be better seen by looking at the equivalent circuit of the 3T configuration, shown in **Figure 22 (a)** for the 3T-IBC tunnel junction (TJ) type and in **Figure 22 (b)** for the 3T-IBC BJT type. Note that due to the extra number of terminals, there is a manifold of possible module wiring configurations for the 3T and 4T configurations, and certain current matching and voltage matching conditions will arise on a module level. Some examples are shown in **Figure 22 (c)-(d)** for 2T, 3T and 4T configurations.

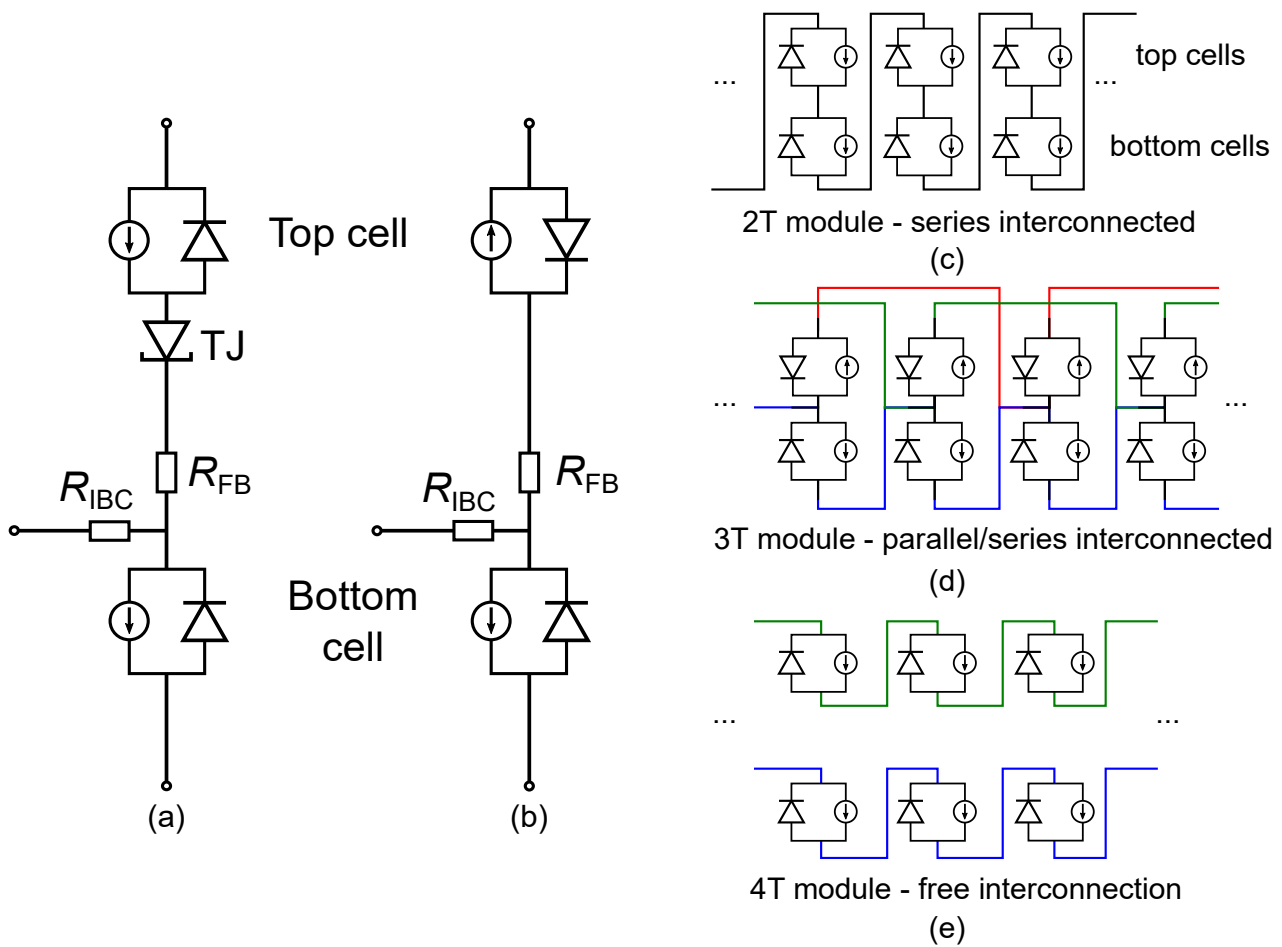

**Figure 22** – Tandem solar cell equivalent circuit for the (a) 3T-IBC with a tunnel junction (TJ) and (b) 3T-IBC in bipolar junction transistor mode; (c) equivalent module circuit for 2T tandems connected in series; (d) one possible equivalent circuit for 3T tandems, with one top cell in series with a parallel of 2 bottom cells (1:2); (e) one possible equivalent circuit for a 4T module, showing a string with the top cells in series and another string with the bottom cells in series. Note: The series and shunt resistors were neglected for simplicity. Figures (a) and (b) are adapted from [145] with permission, copyright © 2019 John Wiley & Sons. Figures (c)-(e) reproduced from [147].

Interestingly, it has been recently shown that a 3T 2J configuration can perform better than a 2T and almost similarly to an independent 4T configuration if the substrings are wired with one top cell in parallel with a series connection of two bottom cells (1:2), as shown in **Figure 22 (d)**. In this case, there is a voltage matching condition that the voltage of the top cell should be similar to the sum of the voltages in the two bottom cells. Incidentally, this condition is matched for ideal combinations of the bandgaps of top and bottom cells, and is approximately matched in real devices (for example, in a GaInP/Si tandem, $V_{oc,GaInP}$ ~ 1.4 V and $V_{oc,Si}$ ~ 0.7 V in ref. [16]). Moreover, this configuration benefits greatly from the fact that the $V_{oc}$ only depends logarithmically on the irradiance, and therefore a voltage matching condition is more resilient against spectral variations than a current matching condition. For that reason, the 3T-IBC configuration is quite promising also on a module level. **Figure 23** summarizes the comparison between 2T, 3T and 4T connections on a cell level and on a module level, as a function of the bandgap of the top cell, for the case where a Si bottom cell is used. Note that in both cases, a parallel connection between the top and bottom cells, either on a cell level (3T-P) or on a module level (1:1 configuration), leads to a flat efficiency curve because the voltage is limited by the bottom cell, and therefore the tandem behaves as a single-junction device.

Another reason why the 3T-IBC structure is considered one of the most promising innovations in multijunction photovoltaics is that it is the only structure, besides the 2T monolithic, that only features one transparent conductive electrode (TCE). This was illustrated above in **Figure 19**. This contrasts with the 4T or 3T-Middle configuration, in which the middle electrodes cause loss of transparency to the bottom cell (this effect also explains the advantage of the 3T configuration over the 4T configuration in **Figure 23 (a)**).

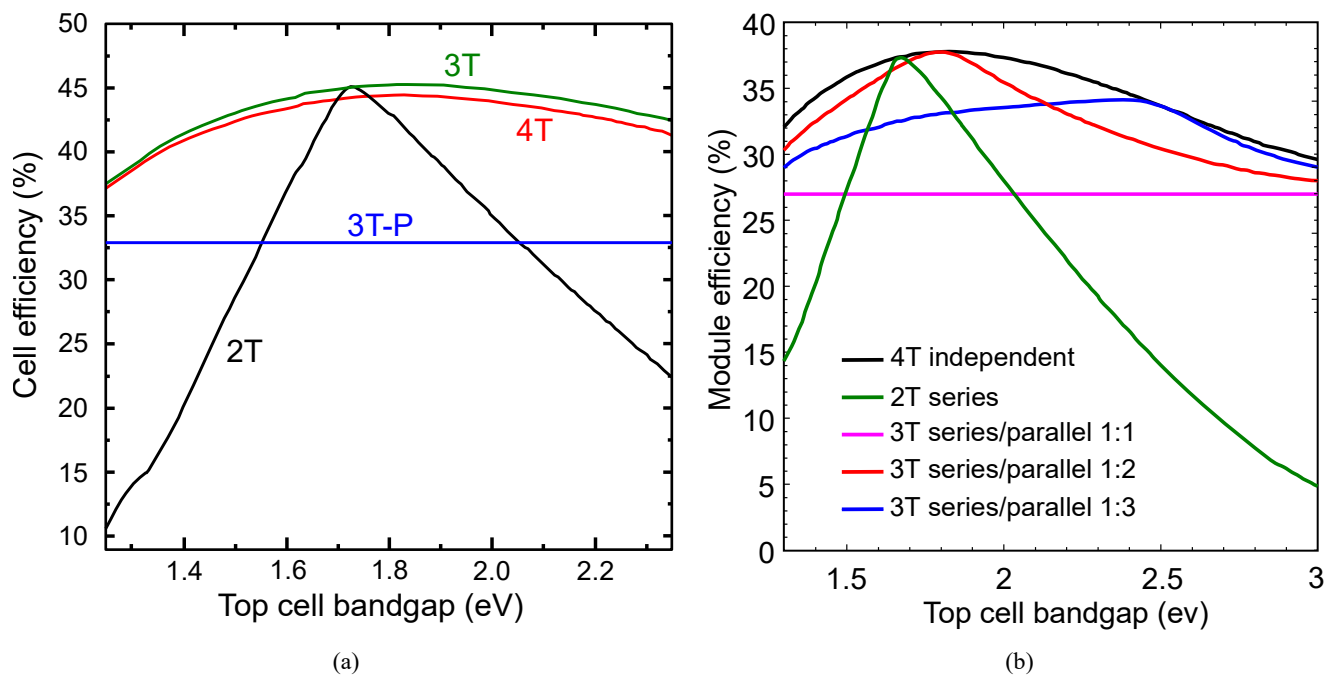

(a)                  (b)

**Figure 23** – (a) Ideal solar cell efficiencies as a function of the top cell bandgap, for different cell configurations, based on Si as bottom cell. Adapted from [148] with permission, copyright © 2020 American Chemical Society; (b) Module efficiency as a function of top cell bandgap for different subcell interconnection configurations, using realistic 2J III-V/Si values. Adapted from [147].

Finally, even though the 3T-IBC structure was originally proposed for the classical current-mismatched InGaP/InGaAS/Ge tandem [141], it is now being intensely researched due to the success of the IBC Si cell, which is the highest efficiency 1J Si architecture at 26.6% efficiency [34]. This high-efficiency IBC structure is achieved with selectively doped a-Si:H contacts patterned on the backside, which can be readily combined with a perovskite top cell due to its low temperature processing. However, this patterned structure could also be formed by diffused junctions or by polysilicon contacts, and allow the fabrication of top cells at higher temperatures.

If these cell and module integration engineering constraints can be solved, higher efficiencies can be achieved by transitioning to multijunction photovoltaics. In the limit where no such constraints exist, the detailed balance efficiency limit for a double-junction tandem can ultimately reach around 45.3%, as illustrated in **Figure 24**. Due to a poorer bandgap matching with the solar spectrum, this value is reduced to 42.5% if Si is used as a bottom cell.

Even then, in that case, this would represent a potential 47% relative increase in efficiency, up from the 29% efficiency limit of 1J Si. Given the market dominance and cost competitiveness of Si, as well as the importance of module efficiency in the overall PV costs, this is potentially a reasonable trade-off. At the same level of module costs, a 50% improvement in module efficiency would directly reduce the price of PV by 33%. Transitioning from a 1J to a 3J Si-based tandem instead, that potential efficiency gain is only 64%, and therefore the initial transition from 1J to 2J devices is also the most impactful. Further increases in the number of junctions give progressively decreasing gains. Ultimately, an infinite number of junctions would yield a 67% efficiency, corresponding to the thermodynamic limit, where the only losses are the Carnot, luminescence emission and solid angle mismatch losses [68]. Given the promising recent developments discussed in this review, this initial transition from 1J to 2J configurations could be realized on a large scale in the coming years, and can potentially be the main driving force for continuous improvements in the PV sector.

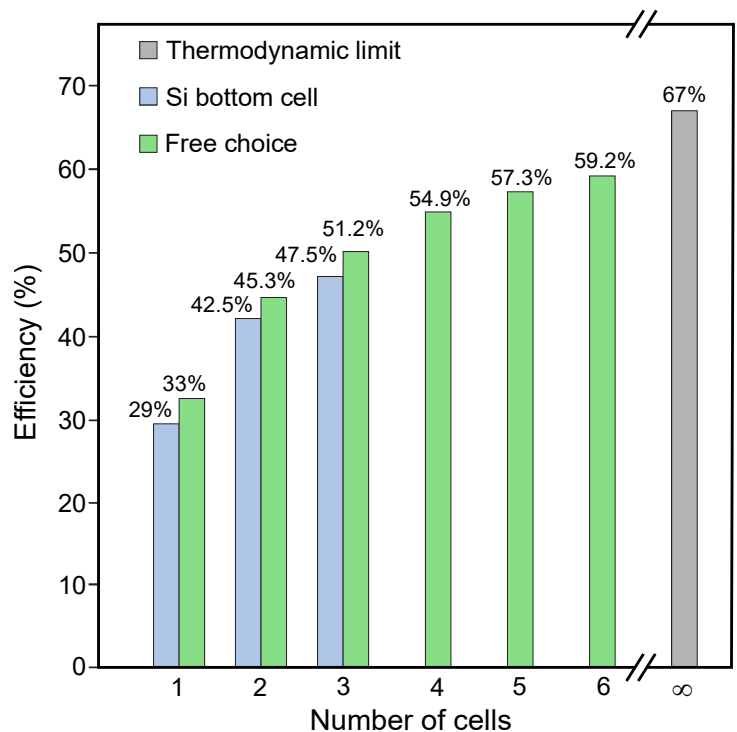

**Figure 24** – (a) Comparison of ideal AM1.5G efficiencies with and without Si as bottom cell. Figure inspired by Green [8].

## 4 Conclusion

Currently, as the uptake of photovoltaics rapidly increases, there are very high expectations from the general public and market makers that the PV sector will continue to further reduce costs and improve the core technology, on path to TW levels of installed capacity. While we do not find any fundamental limitation to expansion in capacity and cost reductions associated with mass production learning curves, the price of PV is now nearly independent of the core cell technology. Instead, the largest fraction of PV costs are area and infrastructure-related, and thus relatively agnostic to the type of solar cells comprising a module, in particular in utility scale applications (upwards of hundreds of kW). Therefore, one of the few remaining ways for further reducing PV costs is to transition to core technologies that enable higher energy yields per unit area. However, the current market technologies are approaching fundamental efficiency limits that will inevitably slow down the rate of technical progress. From a theoretical point of view, the transition from single-junction to multijunction photovoltaics is considered the most promising path to increase the energy density of PV technologies. On the other hand, from a practical point of view, as we discuss in this review work, each different multijunction technology faces specific limitations, such as the cost of the core fabrication processes, insufficient performance or the stability and toxicity of the constituting subcells. As of yet, no specific multijunction device gathers all the aspects necessary for upscaling to enter the PV market on Earth, but there have been considerable recent developments across every category covered in this review work. In all types of III-V tandem cells, increasing manufacturing yields and moving to lower cost deposition methods such as hydride vapor phase epitaxy is crucial to reduce costs. Moreover, decreasing the cost of monocrystalline III-V wafers and/or increasing the wafer reusability is key for the upscaling of metamorphic III-V concepts. In any case, the extremely high efficiency, radiation hardness and temperature stability of these concepts will ensure its continuous development in space applications. Moreover, other high power density applications such as concentration photovoltaics could further bring III-V technologies to the mainstream markets.

The recent developments in new tandem concepts, in particular perovskite/Si, is a major advance in the field, and points to the possibility of new tandem configurations yet to be discovered. Here, compatibility between subcells is an important issue to consider, and the jury is still out regarding the best subcell interconnection configurations. Similarly, there are multiple tandem module integration strategies, and there is no obvious winner at this development stage. Two-terminal monolithic configurations have the obvious advantage of simplicity, but the current matching constraint might limit its applications under variable spectral conditions. On the other hand, three-terminal configurations with interdigitated back contacts could provide outstanding operational flexibility at the module level and leverage the existing knowledge of historically-developed IBC Si structures.

Given these results, we expect future technological improvements in the PV sector to be possible, although such improvements will likely proceed through new concepts such as multijunction configurations, rather than through the incremental improvements in single-junction devices seen over the last decades.

## Acknowledgments

This work was supported by a grant from the Innovation Fund Denmark (Grant 6154-00008A). The author thanks Ole Hansen, Professor Emeritus at the Technical University of Denmark, for insightful suggestions and discussions.

**Table of Contents figure**

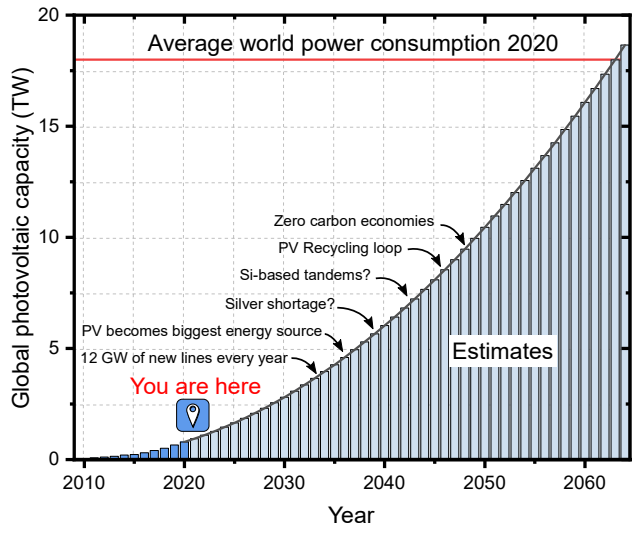